\documentclass[preprint,aps]{revtex4}
\usepackage[dvipdfm]{graphicx}
\def\Vec#1{\mbox{\boldmath $#1$}}

\begin{document}

\title{Monopole Excitation to Cluster States}

\author{
	Taiichi Yamada, 
        Yasuro Funaki$^{1}$,
	Hisashi Horiuchi$^{2,3}$, \\ 
	Kiyomi Ikeda$^{1}$, and Akihiro Tohsaki$^{2}$
}

\affiliation{
Laboratory of Physics, Kanto Gakuin University, Yokohama 236-8501, Japan,} 

\affiliation{
$^{1}$Nishina Center for Accelerator-based Science, 
The Institute of Physical and Chemical Research (RIKEN), Wako 351-0098, 
Japan,}

\affiliation{
$^{2}$Research Center for Nuclear Physics, Osaka University, Osaka 567-0047, 
Japan,}

\affiliation{
$^{3}$International Institute for Advanced Studies, Kizugawa 619-0225, 
Japan}

\date{\today}

\begin{abstract}
We discuss strength of monopole excitation of the ground state to cluster 
states in light nuclei.  We clarify that the monopole excitation to cluster 
states is in general strong  as to be comparable with the single particle 
strength and shares an appreciable portion of the sum rule value in spite 
of large difference of the structure between the cluster state and the 
shell-model-like ground state.  We argue that the essential reasons 
of the large strength are twofold.  One is the fact that the clustering 
degree of freedom is possessed even by simple shell model wave functions.  
The detailed feature of this fact is described by the so-called 
Bayman-Bohr theorem which tells us that  SU(3) shell model wave function 
is equivalent to cluster model wave function.  The other is the ground 
state correlation induced by the activation of the cluster degrees of 
freedom described by the Bayman-Bohr theorem.   We demonstrate, by deriving 
analytical expressions of monopole matrix elements, that the order of 
magnitude of the monopole strength is governed by the first reason, while 
the second reason plays a sufficient role in reproducing the data up to 
the factor of magnitude of the monopole strength.  Our explanation is made 
by analysing three examples which are the monopole excitations to 
the $0^+_2$ and $0^+_3$ states in $^{16}$O and the one to the 
$0^+_2$ state in $^{12}$C.  The present results imply that the measurement 
of strong monopole transitions or excitations is in general very useful 
for the study of cluster states.   
\end{abstract}


\maketitle

\section{Introduction}\label{introduction}

The monopole transitions from cluster states to ground states in light 
nuclei are rather large in comparison with the single particle strength. 
For example in $^{16}$O the monopole matrix elements $M({\rm E0})$ between the 
ground state and the first and second excited $0^+$ states at 
$E_x$ = 6.05 MeV and 12.05 MeV which are known to have $^{12}$C+$\alpha$ 
cluster structure~\cite{supple,hoik,suzuki} are $3.55\pm0.21$~fm$^2$ 
and $4.03\pm0.09$~fm$^2$~\cite{ajzen86}, respectively.  
Also in $^{12}$C the $M({\rm E0})$ value between the ground state and the first 
excited $0^+$ state at $E_x$ = 7.66 MeV (so-called Hoyle state~\cite{Hoyle}) 
which is known to have a $3\alpha$ cluster structure~\cite{supple} is 
$5.4\pm0.2$~fm$^2$~\cite{ajzen86}.  
A rough estimate of the single particle strength 
$\langle u_f(r) | r^2 | u_i(r) \rangle$ is $(3/5)R^2\sim5.4~{\rm fm}^2$ 
for $p$- and $sd$-shell nuclei ($R\sim3.0$ fm).  This estimation formula 
is obtained under the uniform-density approximation of $u(r) 
\sim \sqrt{3/R^3}$ for $u_f(r)$ and $u_i(r)$ with $R$ standing for the 
nuclear radius.  The energy weighted strengths of the above mentioned 
monopole transitions give an appreciable portion of the sum rule values: 
in $^{16}$O they are about 3 $\%$ and 8 \% for $0^+_2$ and $0^+_3$, 
respectively, and in $^{12}$C about 16 \% for $0^+_2$  (see Appendix~\ref{sumrule}). 
Recently Kawabata and his collaborators have studied the excited states  
of $^{11}$B by performing ${^{11}{\rm B}}(d, d')$  reaction and they 
concluded that the third $3/2^-$ state at $E_x$ = 8.56 MeV has a 
$2\alpha + t$ cluster structure~\cite{kawabata}.  Among many reasons for 
this conclusion, one is a large monopole strength for the third $3/2^-$ 
state which is of similar value to the monopole strength for the 
second $0^+$ state in $^{12}$C, and another is that the AMD 
(antisymmetrized molecular dynamics) calculation~\cite{kawabata} 
as well as the $2\alpha+t$ OCM (orthogonality condition model) 
calculation~\cite{yamada07} have reproduced the large monopole strength 
and have assigned loosely bound $2\alpha + t$ cluster structure to the 
third $3/2^-$ state. 

The single particle estimate of the monopole transition is based on 
the assumption that the excited state has a one-particle one-hole excitation 
from the ground state.  However, the cluster structure is very different 
from the shell-model-like structure of the ground state, and its state is 
described as a superposition of many-particle many-hole configurations when 
it is expanded by shell model configurations.   This means that in the 
excited state with a cluster structure, the component of a one-particle 
one-hole excitation from the ground state configuration is expected to be 
very small.   Therefore the observation of rather large monopole strengths 
for cluster states which are comparable with single particle strength 
looks {\it not} to be easy to explain.  The $^{12}$C+$\alpha$ OCM 
calculation~\cite{suzuki} for $^{16}$O and $3\alpha$ RGM 
(resonating group method) calculation~\cite{kami,uega}, however, have 
reproduced rather well the experimental data of the monopole transitions. 
No explicit and detailed analyses of the reason why the cluster models 
reproduce plausibly the experimental data have been presented so far 
as long as we know.  There should exist underlying physics in the 
monopole transition strengths in light nuclei.
 
The purpose of this paper is to clarify the basic reasons why monopole 
transition strength between a cluster state and the ground state in light 
nuclei is generally rather large in comparison with the single particle 
strength and shares an appreciable portion of the sum rule value, 
in spite of the large difference of structure between the initial and 
final states.   We analyse the above-mentioned three cases of monopole 
transisions in $^{16}$O and $^{12}$C, namely the monopole transitions 
between the ground state and the first and second excited $0^+$ states 
in $^{16}$O, and the monopole transition between the ground state and 
the first excited $0^+$ state in $^{12}$C.  By using these analyses  
we will show that there are two basic reasons for the generally large 
strength of monopole transitions.  The first reason is the fact that the 
clustering degree of freedom is possessed even by simple shell model 
wave functions.  The detailed feature of this fact is described by 
the so-called Bayman-Bohr theorem~\cite{BB}.   This theorem tells us 
that the SU(3) shell model wave function~\cite{elliott58} describing 
the ground state is in most cases equivalent to the cluster-model wave 
function discussed by Wildermuth and Kanellopoulos~\cite{Wildermuth}. 
Thus we can see what kinds of clustering degrees of freedom are embedded 
in the ground state. For example the doubly closed-shell wave function 
of the $^{16}$O ground state (total quanta $N_{TOT}=12$) which is just 
the SU(3) shell model wave function with $(\lambda, \mu)=(0, 0)$  
is equivalent to a $^{12}$C + $\alpha$ cluster-model wave function 
with $N_{TOT}=12$.  This means that the ground-state wave function 
of $^{16}$O originally has a $^{12}$C+$\alpha$ clustering degree of 
freedom.  The second reason is the ground state correlation induced 
by the activation of the cluster degrees of freedom described by the 
Bayman-Bohr theorem.  In the case of the above example of the $^{16}$O 
ground state, the ground state correlation is due to the $^{12}$C + 
$\alpha$ clustering degrees of freedom.  As was explained, the first 
and second excited $0^+$ states of $^{16}$O are the cluster states 
with $^{12}$C + $\alpha$ structure.  These cluster states are formed 
just by the excitation of the $^{12}$C + $\alpha$ clustering degree 
of freedom which is already existent in the ground state. Therefore it 
is quite reasonable that the large strength of the monopole transition 
between the ground state and the first and second excited $0^+$ 
states is explained by the above-mentioned first and second reasons. 
We will demonstrate, by deriving analytical expressions of monopole 
matrix elements, that the order of magnitude of the monopole strength 
is governed by the first reason, while the second reason plays a  
sufficient role in reproducing the data up to the factor of 
magnitude of the monopole strength. 

In the present paper we discuss the details of the first and second 
reasons for $^{16}$O and $^{12}$C.   In the case of $^{16}$O, we 
make use of the microscopic $^{12}$C + $\alpha$ cluster wave function, 
while in the case of $^{12}$C, we discuss the problem by using the 
so-called THSR wave function~\cite{tohsaki,funaki}.  Our results 
mean that the measurement of strong monopole transitions provides 
us in general with a very useful tool for the experimental study 
of cluster states as has been practiced in Ref.~\cite{kawabata}. 

The present paper is organized as follows.  In Sec.~\ref{formulation} 
we derive analytical expressions of the monopole matrix elements 
between the ground state and $^{12}$C+$\alpha$ cluster states in 
$^{16}$O and those between the ground state and $3\alpha$ cluster state 
in $^{12}$C, by using the Bayman-Bohr theorem.   
In Sec.~\ref{gr-st-correl} we discuss the effect of the ground 
state correlation on the monopole transitions from the $^{12}$C$ + 
\alpha$ cluster states in $^{16}$O, and those from the $3\alpha$ cluster 
state in $^{12}$C.  In Sec.~\ref{summary} we give discussions and summary.

\section{Monopole Transition and Bayman-Bohr Theorem} \label{formulation}

\subsection{Monopole transition from two-cluster states in $^{16}$O}\label{subsection:Bayman_Bohr}

We discuss the following two observed values of the monopole transition 
matrix  element in $^{16}$O:~One is $M({\rm E0}) = 3.55 \pm 0.21$ fm$^2$ 
between the ground state ($0^+_1$) and the first excited $0^+$ state 
($0^+_2$) at $E_x$ = 6.05 MeV, and  the other is $M({\rm E0}) = 4.03 \pm 0.09$ 
fm$^2$ between the ground state and the excited $0^+$ state ($0^+_3$) 
at $E_x$ = 12.05 MeV.  These excited $0^+$ states are known to have 
$^{12}$C+$\alpha$ structures~\cite{supple,hoik,suzuki}.   
In this section, we explain that the order of magnitude of these $M({\rm E0})$ 
values comparable with the single nucleon strength is explained to 
come from the fact that the doubly closed shell wave function already contains 
in it the $^{12}$C+$\alpha$ clustering degree of freedom. For this purpose  
we derive analytical expressions of these monopole matrix elements by the 
use of the Bayman-Bohr theorem. 

The nuclear SU(3) model or Elliott model~\cite{elliott58} is known to 
describe well ground states of light nuclei.  
The ground state of $^{16}$O has a doubly closed shell structure of 
$0s$ and $0p$ orbits which belongs to the SU(3) irreducible 
representation $(\lambda,\mu)=(0,0)$.  
This doubly closed shell model wave function with the nucleon size 
parameter $\nu_N=M\omega/2\hbar$ ($M$: nucleon mass) 
is equivalent to a cluster wave function of $^{12}$C + $\alpha$ 
configuration, according to the Bayman-Bohr theorem, 
\begin{eqnarray}
  && \frac{1}{\sqrt{16!}} \det |(0s)^4 (0p)^{12} | = N_g \frac{1}
    {\sqrt{{}_{16}C_4}} 
    {\cal A} \{ [{\cal R}_4(\Vec{r},3\nu_N) \phi({}^{12}{\rm C})]_{(0,0)} 
    \phi(\alpha) \}\phi_G(\Vec{r}_G), \label{eq:clshl} \\
  && [{\cal R}_4(\Vec{r},3\nu_N) \phi({}^{12}{\rm C})]_{(\lambda,\mu)=(0,0)} 
    = \sum_{L=0,2,4} C_L \  [{\cal R}_{4L}(\Vec{r},3\nu_N) 
    \phi_L({}^{12}{\rm C})]_{J=0},  \label{eq:alc}  \\
  && \phi_G(\Vec{r}_G)=\left(\frac{32\nu_N}{\pi}\right)^{3/4}
    \exp\left(-16\nu_N\Vec{r}_G^2\right),\ \  \Vec{r}_{\rm G} = 
    \frac{1}{16} \sum_{i=1}^{16} \Vec{r}_i \\
  && C_L =  \langle (4,0)L, (0,4)L ||(0,0)0 \rangle, \ \ {}_{16}C_4 = 
     \frac{16!}{12!4!}.
\end{eqnarray}
Here $\phi(\alpha)$ and $\phi_L({}^{12}{\rm C})$ stand for the internal wave 
 function of $\alpha$ cluster with the $(0s)^4$ configuration and 
 internal wave function of $^{12}$C with angular momentum $L$, respectively. 
$\phi_G$ denotes the center-of-mass wave function of $^{16}$O, which can be 
separated from the internal wave function as is written in 
Eq.~(\ref{eq:clshl}).  The relative wave function between the $\alpha$ and 
$^{12}$C clusters is presented by the harmonic oscillator wave function 
${\cal R}_{NLm}(\Vec{r},\beta) = R_{NL}(r,\beta) Y_{Lm}({\hat{\Vec{r}}})$ 
with the oscillator quanta $N=4$ [nodal number $n = (N-L)/2$] and 
size parameter $\beta=3\nu_N$, where $\Vec{r}$ is the relative coordinate 
between the center-of-masses of $\alpha$ and $^{12}$C clusters. 
It is noted that ${\cal R}_{4L}(\Vec{r},3\nu_N)$ and 
$\phi_L({}^{12}{\rm C})$ belong to the SU(3) irreducible representations 
$(\lambda,\mu)=(4,0)$ and (0,4), respectively.  Equation~(\ref{eq:alc}) 
means that these representations are coupled to the SU(3) scalar 
representation $(\lambda,\mu)=(0,0)$. ${\cal A}$ is the nucleon 
antisymmetrizer between $^{12}$C and $\alpha$ cluster,  $N_g$ is the 
normalization constant, and $C_L$ is the reduced Clebsch-Gordan 
coefficient of SU(3) group for the SU(3) vector coupling $(4,0) \times 
(0,4) \rightarrow (0,0)$.  

The doubly closed shell model wave function of $^{16}$O has the total 
number of the oscillator quanta $N_{TOT}=12$ and is only one possible 
wave function allowed for $N_{TOT}=12$.  Since all three wave functions 
of $^{16}$O,  
${\cal A} \{ [{\cal R}_{4L}(\Vec{r},3\nu_N) \phi_L({}^{12}{\rm C})]_{J=0} 
\phi (\alpha) \}$ for $L$=0, 2, and 4, have the total quanta 
$N_{TOT} = 12$, they are necessarily equivalent to the doubly closed shell 
model wave function $\Phi_{CS}$ and represent the internal wave function 
of the $^{16}$O ground state (see also Appendix~\ref{Baymann_Bohr}), 
\begin{eqnarray}
  |{0^+_1}\rangle &=& \Phi_{CS} \equiv  {\frac{1}{\sqrt{16!}} 
       \det |(0s)^4 (0p)^{12} |} 
              \times [\phi_G(\Vec{r}_G)]^{-1} \label{eq:clshl0} \\ 
    &=& N_{g0}\frac{1}{\sqrt{{}_{16}C_4}} 
        {\cal A} \{ [{\cal R}_{40}(\Vec{r},3\nu_N) 
        \phi_{L=0}({}^{12}{\rm C})]_{J=0} \phi(\alpha) \} \label{eq:clshla} 
         \\ 
    &=& N_{g2} \frac{1}{\sqrt{{}_{16}C_4}} 
         {\cal A} \{ [{\cal R}_{42}(\Vec{r},3\nu_N) 
         \phi_{L=2}({}^{12}{\rm C})]_{J=0} \phi(\alpha) \} \label{eq:clshlb} 
         \\ 
    &=& N_{g4} \frac{1}{\sqrt{{}_{16}C_4}} 
         {\cal A} \{ [{\cal R}_{44}(\Vec{r},3\nu_N) 
         \phi_{L=4}({}^{12}{\rm C})]_{J=0} \phi(\alpha) \},  \label{eq:clshlc} 
\end{eqnarray}
where $N_{g0}$, $N_{g2}$, and $N_{g4}$ denote the normalization constants. 
It is important to recognize the implication of these relations of 
 Eqs.~(\ref{eq:clshl}), and (\ref{eq:clshla})$\sim$(\ref{eq:clshlc}).  
They imply that the ground state of $^{16}$O can 
 be excited not only through single particle degrees of freedom by promoting 
 nucleons from $0s$ and $0p$ orbits to higher orbits, but also through cluster 
 degrees of freedom by exciting the $^{12}$C$-$$\alpha$ relative motion from 
 ${\cal R}_{4L}(\Vec{r},3\nu_N)$ state to higher nodal states. 
The latter characteristic is an essential point to understand
 why the monopole transition matrix elements to cluster states are in general 
 large.  

\subsubsection{Monopole transition between $0_1^+$ and $0_2^+$ states}\label{sec:ME0_16O_21}

The $0_2^+$ state of $^{16}$O is known to have a loosely bound 
$^{12}$C+$\alpha$ structure, in which the dominant component of 
$^{12}$C is the ground state~\cite{supple,hoik,suzuki}.  
Thus we express the $0_2^+$ wave function as 
\begin{equation}
  |0_2^+ \rangle = N_I \frac{1}{\sqrt{{}_{16}C_4}} {\cal A} \{ \chi_0(\Vec{r}) 
     \phi_{L=0}({}^{12}{\rm C}) \phi(\alpha) \},  \label{eq:mysta}
\end{equation}
where $N_I$ represents the normalization constant.  By expanding 
$\chi_0(\Vec{r})$ in terms of harmonic oscillator functions, we have 
\begin{eqnarray}
 |0^+_2 \rangle &=& \sum_{N=6}^\infty \eta_N \Phi_N,  \\ 
  \Phi_N &=& \frac{1}{\sqrt{\tau_{0,N}}} \frac{1}{\sqrt{{}_{16}C_4}} 
    {\cal A} \{ R_{N0}(r,3\nu_N) Y_{00}(\hat{\Vec{r}}) 
    \phi_{L=0}({}^{12}{\rm C}) \phi(\alpha) \}, \\ 
    \tau_{0,N} &\equiv& \langle {\cal R}_{N0}(\Vec{r},3\nu_N) 
  \phi_{L=0}({}^{12}{\rm C}) \phi(\alpha) | {\cal A} 
  \{ {\cal R}_{N0}(\Vec{r},3\nu_N) \phi_{L=0}({}^{12}{\rm C}) 
  \phi(\alpha) \} \rangle.
\end{eqnarray}
It should be noted that $\Phi_N$ are normalized. Also it should be noted 
that $\Phi_{N=4}$ is just the doubly closed shell wave function as is 
seen in Eq.~(\ref{eq:clshla}), $|0_1^+ \rangle = \Phi_{N=4}$.  

Since both $0_1^+$ and $0_2^+$ states have the total isospin $T=0$, 
the monopole transition matrix element $M({\rm E0})$ is 
\begin{eqnarray}
  && M({\rm E0},0_2^+ - 0_1^+) = \langle 0_1^+ | \sum_{i=1}^{16}\frac{1}{2} 
     (1+\tau_{3i}) (\Vec{r}_i - \Vec{r}_G)^2 | 0_2^+ \rangle \nonumber \\ 
  && \ \  = \langle 0_1^+ | \frac{1}{2} \sum_{i=1}^{16} 
     (\Vec{r}_i - \Vec{r}_G)^2 | 0_2^+ \rangle \nonumber \\ 
  && \ \  = \eta_6 \langle \Phi_{N=4} | \frac{1}{2} \sum_{i=1}^{16} 
     (\Vec{r}_i - \Vec{r}_G)^2 | \Phi_{N=6} \rangle. 
\end{eqnarray}
Here $\Vec{r}_G$ stands for the total center-of-mass coordinate, 
$\Vec{r}_{\rm G} = (1/16) \sum_{i=1}^{16} \Vec{r}_i$.  
The last equality is because 
$\sum_{i=1}^{16} (\Vec{r}_i - \Vec{r}_G)^2 \Phi_{N=4}$ can not have 
more than 2$\hbar \omega$ excitation than $\Phi_{N=4}$. Then we have 
\begin{eqnarray}
  && M({\rm E0},0_2^+ - 0_1^+) = \frac{\eta_6}{2} \frac{1}
  {\sqrt{\tau_{0,4}\tau_{0,6}}} \nonumber \\  
  && \ \ \  \times \langle {\cal R}_{40}(\Vec{r},3\nu_N) 
    \phi_{L=0}({}^{12}{\rm C}) \phi(\alpha) | {\cal A} 
     \{ (\sum_{i=1}^{16}(\Vec{r}_i - \Vec{r}_{\rm G})^2) 
     {\cal R}_{60}(\Vec{r},3\nu_N) 
     \phi_{L=0}({}^{12}{\rm C}) \phi(\alpha) \} \rangle \nonumber \\  
  && \ \ = \frac{\eta_6}{2} \frac{1}{\sqrt{\tau_{0,4}\tau_{0,6}}}
     \langle {\cal R}_{40}(\Vec{r},3\nu_N) 
     \phi_{L=0}({}^{12}{\rm C}) \phi(\alpha) | {\cal A} 
     \{ \frac{12 \times 4}{16} \Vec{r}^2 {\cal R}_{60}(\Vec{r},3\nu_N) 
     \phi_{L=0}({}^{12}{\rm C}) \phi(\alpha) \} \rangle.  \label{eq:matr}. 
\end{eqnarray}
In obtaining Eq.~(\ref{eq:matr}), we first used the identity 
\begin{equation}
   \sum_{i=1}^{16}(\Vec{r}_i - \Vec{r}_G)^2 = \sum_{i \in {^{12}{\rm C}} } 
          (\Vec{r}_i - \Vec{r}_C)^2 
    + \sum_{i \in \alpha }(\Vec{r}_i - \Vec{r}_\alpha)^2 +  
    \frac{12 \times 4}{16} \Vec{r}^2,   \label{eq:monop}
\end{equation} 
where $\Vec{r}_{\rm C}$ and $\Vec{r}_\alpha$ express the center-of-mass 
coordinate of $^{12}$C and $\alpha$, respectively.  
We then used the following relations, 
\begin{eqnarray}
   && \langle {\cal R}_{40}(\Vec{r},3\nu_N) \phi_{L=0}({}^{12}{\rm C}) 
      \phi(\alpha) | {\cal A} \{ {\cal R}_{60}(\Vec{r},3\nu_N) \ 
      ( \sum_{i \in {}^{12}{\rm C} } (\Vec{r}_i - \Vec{r}_{\rm C})^2 ) 
      \phi_{L=0}({}^{12}{\rm C})\ \phi(\alpha) \} \rangle  = 0,  
      \label{eq:matra}  \\ 
   && \langle {\cal R}_{40}(\Vec{r},3\nu) \phi_{L=0}({}^{12}{\rm C}) 
      \phi(\alpha) | {\cal A} \{ {\cal R}_{60}(\Vec{r},3\nu_N) \  
      \phi_{L=0}({}^{12}{\rm C})\ (\sum_{i \in \alpha }(\Vec{r}_i 
      - \Vec{r}_\alpha)^2 ) \phi(\alpha) \} \rangle = 0. \label{eq:matrb} 
\end{eqnarray}
These relations can be easily proved by counting the total numbers of 
oscillator quanta of the bra and ket functions. 
First, the number of the oscillator quanta of 
${\cal R}_{60}(\Vec{r},3\nu_N)$ is larger 
than that of ${\cal R}_{40}(\Vec{r},3\nu_N)$ by 2. 
Second, the number of the oscillator quanta of 
$( \sum_i (\Vec{r}_i - \Vec{r}_{\rm C})^2 ) \phi_{L=0}({}^{12}{\rm C})$ 
can not be smaller than that of 
$\phi_{L=0}({}^{12}{\rm C})$ because $\phi_{L=0}({}^{12}{\rm C})$ has the 
smallest number of the oscillator quanta in the $^{12}$C ($N=Z=6$) system.  
Similarly, the number of the oscillator quanta of 
$(\sum_i (\Vec{r}_i - \Vec{r}_\alpha)^2 ) \phi(\alpha)$ 
can not be smaller than that of $\phi(\alpha)$. 
Therefore in each of Eqs.~(\ref{eq:matra}) and (\ref{eq:matrb}), 
the ket function has larger total number of the oscillator 
quanta than that of the bra function at least by 2, which leads to 
the orthogonality of the bra and ket functions. 

Now we expand $(12 \times 4/16) \Vec{r}^2 {\cal R}_{60}(\Vec{r},3\nu_N)$ in 
Eq.~(\ref{eq:matr}) in terms of the harmonic oscillator function  
\begin{equation}
   \frac{12 \times 4}{16} \Vec{r}^2 {\cal R}_{60}(\Vec{r},3\nu_N) = \sum_N 
   \langle {\cal R}_{N0}(\Vec{r},3\nu_N) | \frac{12 \times 4}{16} \Vec{r}^2 | 
   {\cal R}_{60}(\Vec{r},3\nu_N) \rangle \ {\cal R}_{N0}(\Vec{r},3\nu_N) 
   \label{eq:expand}. 
\end{equation}
By inserting Eq.~(\ref{eq:expand}) into Eq.~(\ref{eq:matr}) we obtain 
\begin{eqnarray}
  M({\rm E0},0_2^+ - 0_1^+) &=& \frac{\eta_6}{2} \frac{1}
  {\sqrt{\tau_{0,4}\tau_{0,6}}} \langle 
    {\cal R}_{40}(\Vec{r},3\nu_N) | \frac{12 \times 4}{16} 
    \Vec{r}^2 | {\cal R}_{60}(\Vec{r},3\nu_N) \rangle  \nonumber \\ 
  && \ \ \times  \langle {\cal R}_{40}(\Vec{r},3\nu_N) 
     \phi_{L=0}({}^{12}{\rm C}) \phi(\alpha) | {\cal A} 
     \{ {\cal R}_{40}(\Vec{r},3\nu_N) \phi_{L=0}({}^{12}{\rm C}) 
     \phi(\alpha) \} \rangle  \nonumber \\
  &=& \frac{\eta_6}{2} \sqrt{\frac{\tau_{0,4}}{\tau_{0,6}}} 
    \langle {\cal R}_{40}(\Vec{r},3\nu_N) | 
    \frac{12 \times 4}{16} \Vec{r}^2 | {\cal R}_{60}(\Vec{r},3\nu_N) \rangle  
     \nonumber \\ 
  &=&  \frac{\eta_6}{2} \sqrt{\frac{\tau_{0,4}}{\tau_{0,6}}} 
    \langle R_{40}(r,3\nu_N) | \frac{12 \times 4}{16} r^2 | R_{60}(r,3\nu_N) 
    \rangle.  \label{eq:ezeroa}    
\end{eqnarray}
Here we note the following relation 
\begin{equation}
   \langle R_{40}(r,3\nu_N) | \frac{12 \times 4}{16} r^2 | R_{60}(r,3\nu_N) 
   \rangle 
  = \langle R_{40}(r,\nu_N) | r^2 | R_{60}(r,\nu_N) \rangle, 
\end{equation}
where $R_{N0}(r,\nu_N)$ is the harmonic oscillator radial function 
of single nucleon with the nucleon size parameter $\nu_N$. 
It is noted here that the matrix elements for calculating the single 
 particle E0 matrix element in $^{16}$O are 
 $\langle R_{00}(r,\nu_N) | r^2 | R_{20}(r,\nu_N) \rangle$ and 
 $\langle R_{11}(r,\nu_N) | r^2 | R_{31}(r,\nu_N) \rangle$ 
 which are a few times smaller than the present 
 $\langle R_{40}(r,\nu_N) | r^2 | R_{60}(r,\nu_N) \rangle$ as shown below, 
\begin{eqnarray}
   \langle R_{00}(r,\nu_N) | r^2 | R_{20}(r,\nu_N) \rangle &=&  
      \sqrt{\frac{3}{8}} \frac{1}{\nu_N},  \ \ \   
   \langle R_{11}(r,\nu_N) | r^2 | R_{31}(r,\nu_N) \rangle =  
      \sqrt{\frac{5}{8}} \frac{1}{\nu_N},  \label{mt_of_r2}\\
   \langle R_{40}(r,\nu_N) | r^2 | R_{60}(r,\nu_N) \rangle &=&  
      \sqrt{\frac{21}{8}} \frac{1}{\nu_N}. 
\end{eqnarray}
The reason why the number of oscillator quanta of the relative wave function 
 is higher than those of the single particle wave functions is due to the 
 Fermi statistics of nucleons. 

The final analytical formula of $M({\rm E0},0_2^+ - 0_1^+)$ is expressed as follows, 
\begin{eqnarray}
  M({\rm E0},0_2^+ - 0_1^+) 
   = \frac{1}{2} \sqrt{ \frac{\tau_{0,4}}{\tau_{0,6}}} \eta_6 
       \langle R_{40}(r,\nu_N) | r^2 | R_{60}(r,\nu_N) \rangle. 
       \label{eq:form1}  
\end{eqnarray}

This analytical expression of $M({\rm E0},0_2^+ - 0_1^+)$ 
is our desired result. It explains clearly why $M({\rm E0},0_2^+ - 0_1^+)$ 
has a comparable magnitude as the single nucleon E0 matrix element. 
The factor $\langle R_{40}(r,\nu_N) | r^2 | R_{60}(r,\nu_N) \rangle$ 
is a few times larger than the single nucleon E0 matrix elements 
$\langle R_{00}(r,\nu_N) | r^2 | R_{20}(r,\nu_N) \rangle$ and 
$\langle R_{11}(r,\nu_N) | r^2 | R_{31}(r,\nu_N) \rangle$, while the 
factor $\eta_6$ works to make the E0 value smaller.

\subsubsection{Monopole transition between $0_1^+$ and $0_3^+$ states}

The $0_3^+$ state of $^{16}$O at $E_x$ = 12.05 MeV is known to have also 
 a $^{12}$C + $\alpha$ structure like the $0_2^+$ state~\cite{supple,suzuki}.  
The $^{12}$C cluster in the $0_3^+$ state, however, is not mainly in its 
ground  state like in $0_2^+$ state but dominantly in its excited $2^+$ state 
at $E_x$ = 4.44 MeV.  Thus we can express the $0_3^+$ wave function in a 
good approximation as 
\begin{equation}
  |0_3^+ \rangle = N_{II} \frac{1}{\sqrt{{}_{16}C_4}} {\cal A} 
   \{ [\chi_2(\Vec{r}) \phi_{L=2}({}^{12}{\rm C})]_{J=0} \phi(\alpha) \}. 
\end{equation}
Like in the case of $0_2^+$ state, we expand $\chi_2(\Vec{r})$ in terms of 
harmonic oscillator wave functions and we obtain 
\begin{eqnarray}
 |0^+_3 \rangle &=& \sum_{N=6}^\infty \zeta_N \Psi_N,  \\ 
  \Psi_N &=& \frac{1}{\sqrt{\tau_{2,N}}} \frac{1}{\sqrt{{}_{16}C_4}} 
    {\cal A} \{ [{\cal R}_{N2}(\Vec{r},3\nu_N) 
    \phi_{L=2}({}^{12}{\rm C})]_{J=0} \phi(\alpha) \}, \\ 
  \tau_{2,N} &\equiv& \langle [{\cal R}_{N2}(\Vec{r},3\nu_N) 
  \phi_{L=2}({}^{12}{\rm C})]_{J=0} \phi(\alpha) | {\cal A} 
  \{ [{\cal R}_{N2}(\Vec{r},3\nu_N) \phi_{L=2}({}^{12}{\rm C})]_{J=0} 
  \phi(\alpha) \} \rangle.
\end{eqnarray}
It should be noted that $\Psi_N$ are normalized. Also it should be noted 
that $\Psi_{N=4}$ is just the doubly closed shell wave function as is 
seen in Eq.~(\ref{eq:clshlb}), $|0_1^+ \rangle = \Psi_{N=4}$.  

The calculation of the monopole transition matrix element 
$M({\rm E0},0_3^+ - 0_1^+)$ can be made in the same manner as that of 
$M({\rm E0},0_2^+ - 0_1^+)$ in the previous section, although we use 
Eq.~(\ref{eq:clshlb}) for the $0^+_1$ state of $^{16}$O, 
\begin{eqnarray}
 && M({\rm E0},0_3^+ - 0_1^+)= \langle 0_1^+ |\frac{1}{2} 
 \sum_{i=1}^{16}(\Vec{r}_i - \Vec{r}_{\rm G})^2 | 0_3^+ \rangle \nonumber \\ 
   && \ \  = \frac{1}{2} \sqrt{ \frac{\tau_{2,4}}{\tau_{2,6}}}  
    \zeta_6  
    \langle R_{42}(r,\nu_N) | r^2 | R_{62}(r,\nu_N) \rangle,  \label{eq:form2} 
\end{eqnarray}
where
\begin{eqnarray}
   \langle R_{42}(r,\nu_N) | r^2 | R_{62}(r,\nu_N) \rangle =  
    \frac{3}{2} \frac{1}{\nu_N}.      
\end{eqnarray}

The analytical expression of $M({\rm E0},0_3^+ - 0_1^+)$ in Eq.~(\ref{eq:form2}) 
is our another desired result. Like in the case of $M({\rm E0},0_2^+ - 0_1^+)$, 
it explains clearly why $M({\rm E0},0_3^+ - 0_1^+)$ has also a comparable magnitude 
as the single nucleon E0 matrix element.

\subsubsection{Wave function which absorbs total monopole strength from the 
doubly closed shell} 

The wave function $\Phi_{(2,0)}$ which absorbs total monopole strength from 
the doubly closed shell wave function $\Phi_{CS}$ is given by 
\begin{eqnarray}
  \Phi_{(2,0)} &=& N_{(2,0)} (1 - |\Phi_{CS} \rangle \langle \Phi_{CS}|) O_M 
                 \Phi_{CS}, \label{eq:Phi_20} \\ 
  O_M &=& \frac{1}{2} \sum_{i=1}^{16} (\Vec{r}_i - \Vec{r}_G)^2,
\end{eqnarray}
where $N_{(2,0)}$ is the normalization constant and is presented as
\begin{eqnarray}
  \frac{1}{N_{(2,0)}} & = & \sqrt{ \langle \Phi_{CS}| O_M^2 |\Phi_{CS} \rangle - \langle \Phi_{CS}| O_M |\Phi_{CS} \rangle^2 }
                            \label{eq:normalization_Phi_20}
                        = \sqrt{\frac{69}{32}}\frac{1}{\nu_N}.
\end{eqnarray}
Any wave function $\Phi$ which is orthogonal to both $\Phi_{CS}$ and 
$\Phi_{(2,0)}$ has zero monopole strength from $\Phi_{CS}$, namely 
$\langle \Phi | O_M | \Phi_{CS} \rangle = 0$. This fact is easily derived 
from the orthogonality of $\Phi$ to $\Phi_{CS}$ and $\Phi_{(2,0)}$. 
Then, the monopole strength of the wave function $\Phi_{(2,0)}$ from $\Phi_{CS}$
 is given by 
\begin{eqnarray}
 \langle \Phi_{(2,0)} | O_M | \Phi_{CS} \rangle = \frac{1}{N_{(2,0)}} 
         = \sqrt{\frac{69}{32}}\frac{1}{\nu_N} = \frac{1.47}{\nu_N}. 
 \label{eq:monopole_strength_from_CS}
\end{eqnarray}
Reminding of the relation of 
 $\langle \Phi_{CS}| O_M^2 |\Phi_{CS} \rangle = { \sum_{k} \left|  \langle \Phi_{k} | O_{M} | \Phi_{CS} \rangle \right|^{2} }$
 ($\{ \Phi_{k} \}$ denoting a complete set of wave functions) in Eq.~(\ref{eq:normalization_Phi_20}),
 one finds that the monopole strength in Eq.~(\ref{eq:monopole_strength_from_CS}) corresponds
 to the squared root of the non-energy-weighted sum rule
 ($\sum_{k\not={CS}} |\langle \Phi_{k} | O_{M} | \Phi_{CS} \rangle |^2$)
 of the monopole operator $O_{M}$ with respect to $\Phi_{CS}$,
 i.e.~exhausting the total monopole strength from $\Phi_{CS}$. 

Let us denote by $\Phi_{(2,0)}^{cl}$ the $^{12}$C + $\alpha$ cluster wave 
function which absorbs the total monopole strength from $\Phi_{CS}$ within 
the $^{12}$C + $\alpha$ cluster model space.  $\Phi_{(2,0)}^{cl}$ is not 
equal to $\Phi_{(2,0)}$.  It is because the monopole operator of $^{12}$C 
cluster, $(1/2)\sum_{i \in {^{12}{\rm C}}}(\Vec{r}_i - \Vec{r}_C)^2$,     
and that of the $\alpha$ cluster, $(1/2)\sum_{i \in \alpha}
(\Vec{r}_i - \Vec{r}_\alpha)^2$, which are contained in the total 
monopole operator $O_M$ as seen in Eq.~(\ref{eq:monop}) do excite the 
$^{12}$C and $\alpha$ clusters when $O_M$ operates on $\Phi_{CS}$. 
These excitations of clusters imply that the wave function $\Phi_{(2,0)}$ 
contains components out of the $^{12}$C + $\alpha$ cluster model space. 
The explicit form of $\Phi_{(2,0)}^{cl}$ is given as 
\begin{eqnarray}
 && \Phi_{(2,0)}^{cl} = N_{(2,0)}^{cl} \frac{1}{\sqrt{{}_{16}C_4}} 
    {\cal A} \{ [{\cal R}_6(\Vec{r},3\nu_N) \phi({}^{12}{\rm C})]_{(2,0)} 
    \phi(\alpha) \}, \\ 
 && [{\cal R}_6(\Vec{r},3\nu_N) \phi({}^{12}{\rm C})]_{(2,0)} =  
    \sum_{L=0,2,4} \langle (6,0)L, (0,4)L ||(2,0)0 \rangle  
    [{\cal R}_{6L}(\Vec{r},3\nu_N) \phi_L({}^{12}{\rm C})]_{J=0}. 
\end{eqnarray}
Here $\langle (6,0)L, (0,4)L ||(2,0)0 \rangle$ is the reduced Clebsch-
Gordan coefficient of the SU(3) group for the SU(3) vector coupling 
$(6, 0) \times (0, 4) \rightarrow (2, 0)$. This relation is proved as 
follows. Since the nucleon coordinate $\Vec{r}_i$ is the sum of the 
creation ($\Vec{a}^\dagger_i$) and annihilation($\Vec{a}_i$) operators 
of oscillator quanta, 
$\Vec{r}_i \propto \Vec{a}^\dagger_i + \Vec{a}_i$, 
the monopole operator $O_M$ consists of three parts, 
$O_M = O^{(2,0)}_M + O^{(0,2)}_M + O^{(0,0)}_M$. The number of the 
oscillator quanta is raised by 2 by $O^{(2,0)}_M$, lowered by 2 
by $O^{(0,2)}_M$, and kept unchanged by $O^{(0,0)}_M$. The superfix 
$(\lambda,\mu)$ of the operator $O^{(\lambda,\mu)}_M$ expresses 
its SU(3) tensor character.  Thus the 2$\hbar \omega$-excited 
wave function created by operating $O_M$ on $\Phi_{CS}$ 
necessarily has the SU(3) symmetry (2, 0).  Within the 
$^{12}$C + $\alpha$ cluster model space, $\Phi_{(2,0)}^{cl}$ 
is the only one wave function which is 2$\hbar \omega$-excited and 
has (2, 0) symmetry.  Thus $\Phi_{(2,0)}^{cl}$ absorbs all the 
monopole strength from $\Phi_{CS}$ and other excited wave functions 
orthogonal to $\Phi_{(2,0)}^{cl}$ all have zero monopole strength.   
The monopole strength of  $\Phi_{(2,0)}^{cl}$ is given by (see Ref.~\cite{hecht}) 
\begin{eqnarray}
  \langle \Phi_{(2,0)}^{cl} | O_M | \Phi_{CS} \rangle 
         = \sqrt{\frac{45}{32}}\frac{1}{\nu_N} = \frac{1.19}{\nu_N}. 
\end{eqnarray}
This magnitude of $\langle \Phi_{(2,0)}^{cl} | O_M | \Phi_{CS} 
\rangle$ is about 80~\% of the total monopole strength 
$\langle \Phi_{(2,0)} | O_M | \Phi_{CS} \rangle$.  We now know, 
from the studies in previous subsections, that the reason of this 
large value is just because of the $^{12}$C + $\alpha$ clustering 
character embedded in the doubly closed shell wave function 
which is described by the Bayman-Bohr theorem.  Namely, 
$\langle \Phi_{(2,0)}^{cl} | O_M | \Phi_{CS} \rangle$ can be 
expressed as 
\begin{eqnarray}
  && \langle \Phi_{(2,0)}^{cl} | O_M | \Phi_{CS} \rangle 
   = \sum_{L=0,2,4} E_L  
   \frac{1}{2} \sqrt{\frac{\tau_{L,4}}{\tau_{L,6}}} 
   \langle R_{4L}(r,\nu_N) | r^2 | R_{6L}(r,\nu_N) \rangle, \\ 
  && \frac{1}{2} \sqrt{ \frac{\tau_{0,4}}{\tau_{0,6}}} 
    \langle R_{40}(r,\nu_N) | r^2 | R_{60}(r,\nu_N) \rangle  
    = \frac{0.784}{\nu_N}, \label{eq:mono0} \\
  && \frac{1}{2} \sqrt{ \frac{\tau_{2,4}}{\tau_{2,6}}} 
    \langle R_{42}(r,\nu_N) | r^2 | R_{62}(r,\nu_N) \rangle  
    = \frac{1.03}{\nu_N}, \label{eq:mono2} \\
  && \frac{1}{2} \sqrt{ \frac{\tau_{4,4}}{\tau_{4,6}}} 
    \langle R_{44}(r,\nu_N) | r^2 | R_{64}(r,\nu_N) \rangle  
    = \frac{0.946}{\nu_N}. \label{eq:mono4}
\end{eqnarray}
The definition of $\tau_{L,N}$ which is already given for 
$L=$ 0 and 2 is as follows 
\begin{eqnarray}
\tau_{L,N} &\equiv& \langle [{\cal R}_{NL}(\Vec{r},3\nu_N) 
  \phi_L({}^{12}{\rm C})]_{J=0} \phi(\alpha) | {\cal A} 
  \{ [{\cal R}_{NL}(\Vec{r},3\nu_N) \phi_L({}^{12}{\rm C})]_{J=0} 
  \phi(\alpha) \} \rangle.
\end{eqnarray}
The coefficient $E_L$ is expressed as follows 
\begin{eqnarray}
 E_L &=& N_{(2,0)}^{cl} \sqrt{\tau_{L,6}} \langle (6,0)L, (0,4)L ||
     (2,0)0 \rangle,  \\
 \left(\frac{1}{N_{(2,0)}^{cl}}\right)^2 &=& 
  \langle [{\cal R}_6(\Vec{r},3\nu_N) \phi({}^{12}{\rm C})]_{(2,0)} 
  \phi(\alpha) | {\cal A} \{ [{\cal R}_6(\Vec{r},3\nu_N) 
  \phi({}^{12}{\rm C})]_{(2,0)} \phi(\alpha) \} \rangle \\ 
 &=& \frac{112}{81} = 1.38.      
\end{eqnarray}
The values of $\langle (6,0)L, (0,4)L ||(2,0)0 \rangle$ are 
$\sqrt{1/10}$, $\sqrt{3/7}$, and $\sqrt{33/70}$, for $L$ = 0, 2, 
and 4, respectively. 
The value of $(1/N_{(2,0)}^{cl})^2$ is given in Ref.~\cite{horic} 
with the notation $\mu^6_{(2,0)}$.  
In Eqs.~(\ref{eq:mono0}) $\sim$ (\ref{eq:mono4}) we used the values of 
$\tau_{L,N}$ calculated by the use of their analytical 
expressions presented in Refs.~\cite{suzuki,horic}.  The values of 
$\tau_{0,N}$ and $\tau_{2,N}$ are given in Table I. 
As we already emphasized, each term  
$(1/2) \sqrt{\tau_{L,4}/\tau_{L,6}} 
\langle R_{4L}(r,\nu_N) | r^2 | R_{6L}(r,\nu_N) \rangle$ 
is all large comparable with single nucleon strength. 

The monopole matrix elements between the $0_2^+$ and $0_3^+$ 
states and the $0_1^+$ state can be calculated by using 
$\Phi_{(2,0)}^{cl}$ as follows 
\begin{eqnarray}
  && M({\rm E0},0_2^+ - 0_1^+) = \langle 0_2^+ | 
  \Phi_{(2,0)}^{cl} \rangle 
  \langle \Phi_{(2,0)}^{cl} | O_M | \Phi_{CS} \rangle 
  = \langle 0_2^+ | \Phi_{(2,0)}^{cl} \rangle \sqrt{\frac{45}{32}} 
  \frac{1}{\nu_N}, \label{eq:SmonoI} \\
  && M({\rm E0},0_3^+ - 0_1^+) = \langle 0_3^+ | 
  \Phi_{(2,0)}^{cl} \rangle 
  \langle \Phi_{(2,0)}^{cl} | O_M | \Phi_{CS} \rangle 
  = \langle 0_3^+ | \Phi_{(2,0)}^{cl} \rangle \sqrt{\frac{45}{32}} 
  \frac{1}{\nu_N}. \label{eq:SmonoII}
\end{eqnarray}

\subsection{Monopole transition from three-cluster state in $^{12}$C}

The calculation of the monopole transition from three-cluster state in 
$^{12}$C can be made essentially in the same way as in the case of 
two-cluster state. We explain this point by calculating the monopole 
transition matrix element from the second $0^+$ state at $E_x = 7.66$ MeV 
to the ground state.  
The experimental data is $M({\rm E0},0_2^+ - 0_1^+)=5.4\pm0.2$ fm$^{2}$.
In the previous section we described the ground state ($0^+_1$) of 
$^{12}$C by the SU(3) shell model wave function 
$\phi_{L=0}({}^{12} {\rm C})$ which belongs to 
the SU(3) irreducible representation $(\lambda,\mu)$ = (0,4). 
This wave function is known of course to be a rather good approximation.  
According to the Bayman-Bohr theorem the internal wave function of 
 the $^{12}$C ground state can be expressed in terms of 
 the $3\alpha$ cluster wave function, 
\begin{eqnarray}
  |0_1^+ \rangle &=& {|(0s)^4 (0p)^8 (0,4) J=0 \rangle}_{internal} \nonumber\\
                 &=& {\widehat N}_g 
  \sqrt{\frac{4!4!4!}{12!}} {\cal A} \{ {\widehat g}_{(04)0}
  (\Vec{s},\Vec{t}) \phi(\alpha_1)\phi(\alpha_2)\phi(\alpha_3) \}, 
\end{eqnarray}  
 where $\Vec{s}$ and $\Vec{t}$ are the Jacobi coordinates defined by 
\begin{equation}
  \Vec{s} = \Vec{X}_2 - \Vec{X}_1,\ \ \ \Vec{t} = \Vec{X}_3 - 
  \frac{\Vec{X}_2 + \Vec{X}_1}{2}, \ \ \ 
  \Vec{X}_k = \frac{1}{4} \sum_{i \in \alpha_k} \Vec{r}_i,  
\end{equation}
 and ${\cal A}$ is antisymmetrizer among nucleons belonging to different 
 $\alpha$ clusters.  
The relative wave function ${\widehat g}_{(04)0}(\Vec{s},\Vec{t})$ is 
expressed as follows 
\begin{eqnarray}
   {\widehat g}_{(04)0}(\Vec{s},\Vec{t}) &=& \sum_{L=0,2,4} 
   \langle (4,0)L, (4,0)L || 
   (0,4) J=0 \rangle {\cal R}_{4,4,L}^{8,J=0}(\Vec{s},\Vec{t}), \\
   {\cal R}^{N,J=0}_{N_1,N_2,L}(\Vec{s},\Vec{t}) &\equiv& 
   [{\cal R}_{N_1 L}(\Vec{s},2\nu_N) 
     {\cal R}_{N_2 L}(\Vec{t},\frac{8}{3}\nu_N)]_{J=0},  \ \  (N_1 + N_2 = N). 
      \label{eq:expb}
\end{eqnarray}
 where ${\cal R}_{NL}(\Vec{u},\beta)$ stands for the harmonic oscillator 
 function of the size parameter 
 $\beta$ of the coordinate $\Vec{u}$ with the oscillator quantum number 
 $N$ and angular momentum $L$.

The SU(3) symmetry $(0,4)$ for $(0s)^4(0p)^8$ configuration is equivalent 
to the spatial symmetry $[44]$ for $(0s)^4(0p)^8$ configuration.  
Since there is only one state with $J=0$
 for the $(0s)^4(0p)^8 [44]$ configuration, the following identities hold 
 (see also Appendix~\ref{Baymann_Bohr}), 
\begin{eqnarray}
  |0_1^+ \rangle &=& {\widehat N}_{g0} \sqrt{\frac{4!4!4!}{12!}} {\cal A} 
   \{ {\cal R}_{4,4,L=0}^{8,J=0}(\Vec{s},\Vec{t})  
   \phi(\alpha_1)\phi(\alpha_2)\phi(\alpha_3) \}  \label{eq:c12a}  \\ 
  &=& {\widehat N}_{g2} \sqrt{\frac{4!4!4!}{12!}} {\cal A} 
   \{ {\cal R}_{4,4,L=2}^{8,J=0}(\Vec{s},\Vec{t}) 
   \phi(\alpha_1)\phi(\alpha_2)\phi(\alpha_3) \}   \label{eq:c12b}  \\ 
  &=& {\widehat N}_{g4} \sqrt{\frac{4!4!4!}{12!}} {\cal A} 
   \{ {\cal R}_{4,4,L=4}^{8,J=0}(\Vec{s},\Vec{t}) 
   \phi(\alpha_1)\phi(\alpha_2)\phi(\alpha_3) \}.  \label{eq:c12c} 
\end{eqnarray}

\subsubsection{Monopole transition between the ground and Hoyle states}

The second $0^+$ state ($0^+_2$) is known to have $3\alpha$ structure 
\cite{supple}, and so we express its wave function as 
\begin{equation}
  |0_2^+ \rangle = {\widehat N}_H \sqrt{\frac{4!4!4!}{12!}} {\cal A} 
  \{ {\widehat \chi}_H(\Vec{s},\Vec{t}) \phi(\alpha_1) \phi(\alpha_2) 
  \phi(\alpha_3) \}. 
\end{equation}
In the expansion of the relative wave function 
${\widehat \chi}_H(\Vec{s},\Vec{t})$ 
 in terms of the harmonic oscillator wave functions, the number of the 
 total oscillator quanta of these oscillator wave functions is larger 
 than 8 which is the number of total oscillator quanta of relative wave 
 function of the ground state. 
Just in the same manner as in the previous section, we can express 
 the monopole transition matrix element $M({\rm E0},0_2^+ - 0_1^+)$ 
 as follows, 
\begin{eqnarray}
  && M({\rm E0},0_2^+ - 0_1^+) = \langle 0_1^+ | \frac{1}{2} \sum_{i=1}^{12}
     (\Vec{r}_i - \Vec{r}_G)^2 | 0_2^+ \rangle \nonumber \\ 
  && \ = \frac{1}{2} \sum_L \frac{{\widehat N}_H}
     {{\widehat N}_{gL}} 
     \langle {\cal R}^{8,J=0}_{4,4,L}(\Vec{s},\Vec{t}) | 
     ( 2 \Vec{s}^2 + \frac{8}{3} \Vec{t}^2 ) | 
     {\widehat \chi}_H(\Vec{s},\Vec{t}) \rangle.   \label{eq:ezeroc}
\end{eqnarray}
Here we used the following relation,
\begin{eqnarray}
 \sum_{i=1}^{12}(\Vec{r}_i - \Vec{r}_G)^2 = \sum_{k=1}^3  
    \sum_{i\in\alpha_k} (\Vec{r}_i - {\bf X}_k)^2 + 2 \Vec{s}^2
    + \frac{8}{3}\Vec{t}^2.   \label{eq:sqrms_12C} 
\end{eqnarray}
It is noted that the first term in Eq.~(\ref{eq:sqrms_12C}) does not 
 contribute to the monopole transition matrix element like in the 
 previous case of $^{16}$O. 

The second $0^+$ state in $^{12}$C which is known as the Hoyle state  
 has been studied by many authors with 3$\alpha$ cluster model and its 
 structure is now regarded as being mainly composed of weakly 
 interacting $3\alpha$ clusters mutually in $S$-wave~\cite{supple,hori,kami,uega}.  
Therefore we write ${\widehat \chi}_H(\Vec{s},\Vec{t})$ as follows 
\begin{equation}
  {\widehat \chi}_H(\Vec{s},\Vec{t}) = {\widetilde \chi}_H(s,t)
    Y_{00}(\Vec{\hat s}) Y_{00}(\Vec{\hat t}). 
\end{equation}
As we already mentioned, the expansion ${\widetilde \chi}_H(s,t)$ 
 in terms of the harmonic oscillator function does not contain 
 the components whose numbers of oscillator quanta are less than or 
 equal to 8, 
\begin{eqnarray}
   {\widetilde \chi}_H(s,t) &=& \sum_{N_1,N_2} D_{H,N_1,N_2} 
   R_{N_1,0}(s,2\nu_N)
   R_{N_2,0}(t,\frac{8}{3}\nu_N), \label{eq:hoexp}  \\  
   D_{H,N_1,N_2} &=& 0 \ \ \  {\rm for} \ \  N_1+N_2 \le 8.  
\end{eqnarray}
Substituting Eq.~(\ref{eq:hoexp}) into Eq.~(\ref{eq:ezeroc}), we have 
 the following simple result 
\begin{equation}
  M({\rm E0},0_2^+ - 0_1^+) = \frac{1}{2} \frac{{\widehat N}_H}{{\widehat N}_{g0}} 
  ( D_{H,6,4} + D_{H,4,6} ) \langle R_{40}(r,\nu_N) | r^2 | R_{60}(r,\nu_N) 
  \rangle. 
  \label{eq:monopole_Hoyle}
\end{equation}
This analytical expression explains why $M({\rm E0},0_2^+ - 0_1^+)$ 
has a comparable magnitude as the single nucleon E0 matrix element. 
The factor $\langle R_{40}(r,\nu_N) | r^2 | R_{60}(r,\nu_N) \rangle$, 
which appears also in the case of $^{16}$O,  
is a few times larger than the single nucleon E0 matrix elements 
$\langle R_{00}(r,\nu_N) | r^2 | R_{20}(r,\nu_N) \rangle$ and 
$\langle R_{11}(r,\nu_N) | r^2 | R_{31}(r,\nu_N) \rangle$, while the 
other factors work to make the E0 value smaller.  The reason why 
we have this formula of Eq.~(\ref{eq:monopole_Hoyle}) is just the 
3$\alpha$ clustering character of the shell model wave function 
of the $^{12}$C ground state which is described by the Bayman-Bohr 
theorem.

\subsubsection{Description of the Hoyle state as a 3$\alpha$ condensate}
\label{subsection:hoyle_state}

Recently the structure of the Hoyle state has been studied from a new 
 point of view that this state is the Bose-condensed state of $3\alpha$ 
 particles~\cite{tohsaki,yamada,funaki}.  
It has been demonstrated that both of the 3$\alpha$ wave functions of 
 Refs.~\cite{kami} and \cite{uega} which are the full solutions of 
 3$\alpha$ Resonating Group Method (RGM) equation of motion have large 
 overlaps close to 100 \% 
 with the 3$\alpha$ Bose-condensed wave functions~\cite{funaki}.  
Therefore we here adopt as ${\widehat \chi}_H(\Vec{s},\Vec{t})$ the 
following form 
\begin{eqnarray}
  {\widehat \chi}_H(\Vec{s},\Vec{t}) &=& ( 1 - P )\  
    {\widehat \chi}_{HG}(\Vec{s},\Vec{t}), \\ 
  {\widehat \chi}_{HG}(\Vec{s},\Vec{t}) &\equiv& 
    \left( \frac{8\gamma}{\sqrt{3} \pi} \right)^{\frac{3}{2}} 
    \exp \{ -4\gamma \sum_{k=1}^3 ({\bf X}_k - \Vec{r}_G)^2 \} \nonumber \\ 
  &=& \left( \frac{4\gamma}{\pi} \right)^{\frac{3}{4}}  
    \left(\frac{16\gamma}{3\pi}\right)^{\frac{3}{4}} 
    \exp \{ - \gamma ( 2\Vec{s}^2 + \frac{8}{3} \Vec{t}^2 ) \}, \\ 
   P &\equiv& \sum_{N \le 8} \sum_{N_1+N_2=N} \sum_L 
    |{\cal R}^{N,J=0}_{N_1,N_2,L}(\Vec{s},\Vec{t}) \rangle  \langle 
     {\cal R}^{N,J=0}_{N_1,N_2,L}(\Vec{s},\Vec{t})|,
\end{eqnarray}
 where $\gamma$ denotes the width parameter which characterizes the 
 $3\alpha$ condensate wave function. 
$P$ is the projection operator onto the state of SU(3) relative motion 
of the ground state and the states forbidden by the antisymmetrization. 
Then, the analytical expression of the monopole transition matrix element 
 in Eq.~(\ref{eq:monopole_Hoyle}) is given as follows: 
\begin{eqnarray}
  M({\rm E0},0_2^+ - 0_1^+) &=& \sqrt{\frac{7}{6}} \sqrt{\frac{\langle F_4 \rangle} 
    {\langle F_5 \rangle}} \xi_5 \langle R_{40}(r,\nu_N) | r^2 | 
    R_{60}(r,\nu_N) \rangle,  \label{eq:form3}  \\
  \xi_5 &\equiv& \sqrt{\frac{\langle F_5 \rangle}{\langle F_5 \rangle + 
    \sum_{n=6}^\infty \displaystyle{ \left( \frac{\nu_N-\gamma}
    {\nu_N+\gamma} \right)^{2(n-5)} }  
    \langle F_n \rangle}}, 
\end{eqnarray}
where the definitions of $F_n$ and $\langle F_n \rangle$ are 
\begin{eqnarray}
  && {\widehat \chi}_{H}(\Vec{s},\Vec{t}) = \left( \frac{2\sqrt{\nu_N\gamma}} 
    {\nu_N+\gamma} \right)^3 \sum_{n=5}^\infty \left( \frac{\nu_N-\gamma} 
    {\nu_N+\gamma} \right)^n  F_n(\Vec{s},\Vec{t}), \\ 
  &&  F_n(\Vec{s},\Vec{t}) = \sum_{n_1+n_2=n} 
    \sqrt{\frac{(2n_1+1)!!(2n_2+1)!!} {(2n_1)!!(2n_2)!!}} 
    {\cal R}^{2n,J=0}_{{2n_1},{2n_2},L=0}(\Vec{s},\Vec{t}), \\    
  && \langle F_n \rangle = \langle F_n(\Vec{s},\Vec{t}) \phi(\alpha_1) 
     \phi(\alpha_2) \phi(\alpha_3) | {\cal A} \{ F_n(\Vec{s},\Vec{t}) 
     \phi(\alpha_1) \phi(\alpha_2) \phi(\alpha_3) \} \rangle. 
\end{eqnarray}
We see that the dependence of $M({\rm E0},0_2^+ - 0_1^+)$ on the parameter 
 $\gamma$ is contained only in the factor $\xi_5$. 
The derivation of the above analytical expression of $M({\rm E0},0_2^+ - 0_1^+)$ 
 is given in Appendix~\ref{Hoyle}.

\section{Ground State Correlations}\label{gr-st-correl}

We first study in this section the numerical values of the monopole 
matrix elements in $^{16}$O and $^{12}$C, by using the formulae 
obtained in the previous section.  As is expected from the analytical 
forms, the numerical values are shown to have the same order of 
magnitude as the observed values which are comparable with the 
single nucleon strength.  However, the calculated values are 
found to be smaller by a few times in $^{16}$O and by several 
times in $^{12}$C.  Therefore we next study in this section 
the effect of the ground state correlation on the magnitude 
of the monopole matrix elements. We will see that the ground state 
correlation largely improves the reproduction of the observed values 
up to the factor of magnitude.  The ground state correlation we consider 
is due to the activation of the clustering degree of freedom which 
is described by the Bayman-Bohr theorem.

\subsection{Monopole transition matrix elements in $^{16}$O}\label{results:16O}

We calculate $M({\rm E0},0_2^+ - 0_1^+)$ and $M({\rm E0},0_3^+ - 0_1^+)$ in $^{16}$O by 
using their analytical expressions given in Eqs.~(\ref{eq:form1}) and 
(\ref{eq:form2}).  The nucleon size parameter $\nu_N$ 
($\nu_N=0.151$ fm$^{-2}$) is chosen so as to reproduce the experimental rms 
radius of $^{16}$O whose wave function is described with the doubly closed 
shell configuration $\Phi_{CS}$.  We also use the expressions given in 
Eq.~(\ref{eq:mono0}) and (\ref{eq:mono2}).  The realistic values of 
$\eta_6$ and $\zeta_6$, of course, should be obtained from the structure 
calculation. A representative structure calculation is that of 
Ref.~\cite{suzuki} in which the $^{12}$C + $\alpha$ OCM was adopted. For 
the sake of the study of this paper, we repeated the same calculation as 
Ref.~\cite{suzuki}.  According to the results, the $0^+_2$ wave function of 
$^{16}$O has the predominant component of $^{12}$C($0^+$)+$\alpha$ channel 
and small components of $^{12}$C($2^+$)+$\alpha$ and $^{12}$C($4^+$)+$\alpha$ 
channels, as mentioned already. Similarly, the $0^+_3$ wave function of 
$^{16}$O has the large component of $^{12}$C($2^+$)+$\alpha$ channel and small 
components of $^{12}$C($0^+$)+$\alpha$ and $^{12}$C($4^+$)+$\alpha$ channels. 

By using the results in Table I of the first paper of Ref.~\cite{suzuki}, we 
get approximate values for $\eta_6$ and $\zeta_6$, and they are 
$\sqrt{0.144}=0.379$ and $\sqrt{0.325}=0.570$, respectively.  We explain 
the details in Appendix~\ref{suzuki}.  Then, the monopole matrix elements 
are estimated to be
\begin{eqnarray}
  M({\rm E0},0_2^+ - 0_1^+) &=& 0.379 \times \frac{0.784}{\nu_N} = 1.97~{\rm fm}^2, 
   \label{eq:result1}  \\
  M({\rm E0},0_3^+ - 0_1^+) &=& 0.570 \times \frac{1.03}{\nu_N} = 3.88~{\rm fm}^2,
   \label{eq:result2}
\end{eqnarray}
where the formulae in Eqs.~(\ref{eq:mono0}) and (\ref{eq:mono2}) are used. 
The result for $M({\rm E0},0_2^+ - 0_1^+)$ gives about 60~\% of the experimental 
value, and the result for $M({\rm E0},0_3^+ - 0_1^+)$ amounts to 96~\% of the 
experimental one. 

In estimating the above values for $M({\rm E0},0_2^+ - 0_1^+)$ and $M({\rm E0},0_3^+ - 0_1^+)$,
 we used Eqs.~(\ref{eq:form1}) and (\ref{eq:form2}), respectively. 
These formulae are based on the assumption that the wave functions of the 
$0_2^+$ and $0_3^+$ states are purely of the structure of $^{12}$C($0^+$) + 
$\alpha$ and $^{12}$C($2^+$) + $\alpha$, respectively. However, of course, 
the OCM calculation of Ref.~\cite{suzuki} shows the small contamination 
of other channel components than the dominant component. So we here also 
give values which take into account these small contamination of other 
non-dominant channel components. Such values can be obtained by using the 
formulae given in Eqs.~(\ref{eq:SmonoI}) and (\ref{eq:SmonoII}). 
The values $\langle 0_2^+ | \Phi_{(2,0)}^{cl} \rangle $ and 
$\langle 0_3^+ | \Phi_{(2,0)}^{cl} \rangle $ are given in Table I of the 
second paper of Ref.~\cite{suzuki}, and they are 0.171 and 0.321, 
respectively.  Then, the monopole matrix elements are estimated to be
\begin{eqnarray}
  M({\rm E0},0_2^+ - 0_1^+) &=& 0.171 \times \frac{1.19}{\nu_N} = 1.35~{\rm fm}^2, 
   \label{eq:result3} \\
  M({\rm E0},0_3^+ - 0_1^+) &=& 0.321 \times \frac{1.19}{\nu_N} = 2.53~{\rm fm}^2. 
   \label{eq:result4}
\end{eqnarray}
Now this result for $M({\rm E0},0_2^+ - 0_1^+)$ gives about 38~\% of the 
experimental value, while the result for $M({\rm E0},0_3^+ - 0_1^+)$ gives about 
63~\% of the experimental one.  These values are smaller than the former 
values in Eqs.~(\ref{eq:result1}) and (\ref{eq:result2}), but still they 
are different from the observed values only by a few factors.  The reason 
why the former values in Eqs.~(\ref{eq:result1}) and (\ref{eq:result2}) are 
larger than the corresponding present values is that the magnitudes of the 
$\Phi_{(2,0)}^{cl}$ component in the pure channel wave functions, 
$^{12}$C($0^+$) + $\alpha$ for $0_2^+$ and $^{12}$C($2^+$) + $\alpha$ for 
$0_3^+$, are larger than those of the OCM wave functions for the $0_2^+$ 
and $0_3^+$ states, respectively.  We give detailed explanation in the 
Appendix~\ref{suzuki}.

As we see above the calculated values of the monopole matrix elements 
surely reproduce the order of magnitude of the observed values which are 
comparable with single nucleon strength. But when compared with the 
data in detail, they are a few times smaller than the observed values. 
We below show that when we take into account the ground state correlation 
theoretical values are improved as to attain the reproduction of the data 
within 10 $\sim$ 20 \% accuracy.  

The ground state correlation we consider is the one caused by the activation 
of the clustering degree of freedom described by Bayman-Bohr theorem. 
In previous sections we demonstrated that the clustering degree of freedom 
described by Bayman-Bohr theorem is the very reason why the monopole 
strengths of excited cluster states are so large as to be comparable with 
single nucleon strength.  However, we only considered the clustering degree 
of freedom rather in a static way. Namely we did not consider the dynamical 
effect of the clustering degree of freedom which excites the ground state 
configuration toward including higher quantum configurations. We know that 
the clustering degree of freedom described by Bayman-Bohr theorem has the 
physical reality because we observe many excited cluster states which are 
formed by exciting the clustering degree of freedom embedded in the ground 
state. Therefore taking into account the ground state correlation caused 
by the clustering degree of freedom described by Bayman-Bohr theorem is 
very natural and should be studied.  

\begin{figure}
\begin{center}
\includegraphics*[scale=0.5,clip]{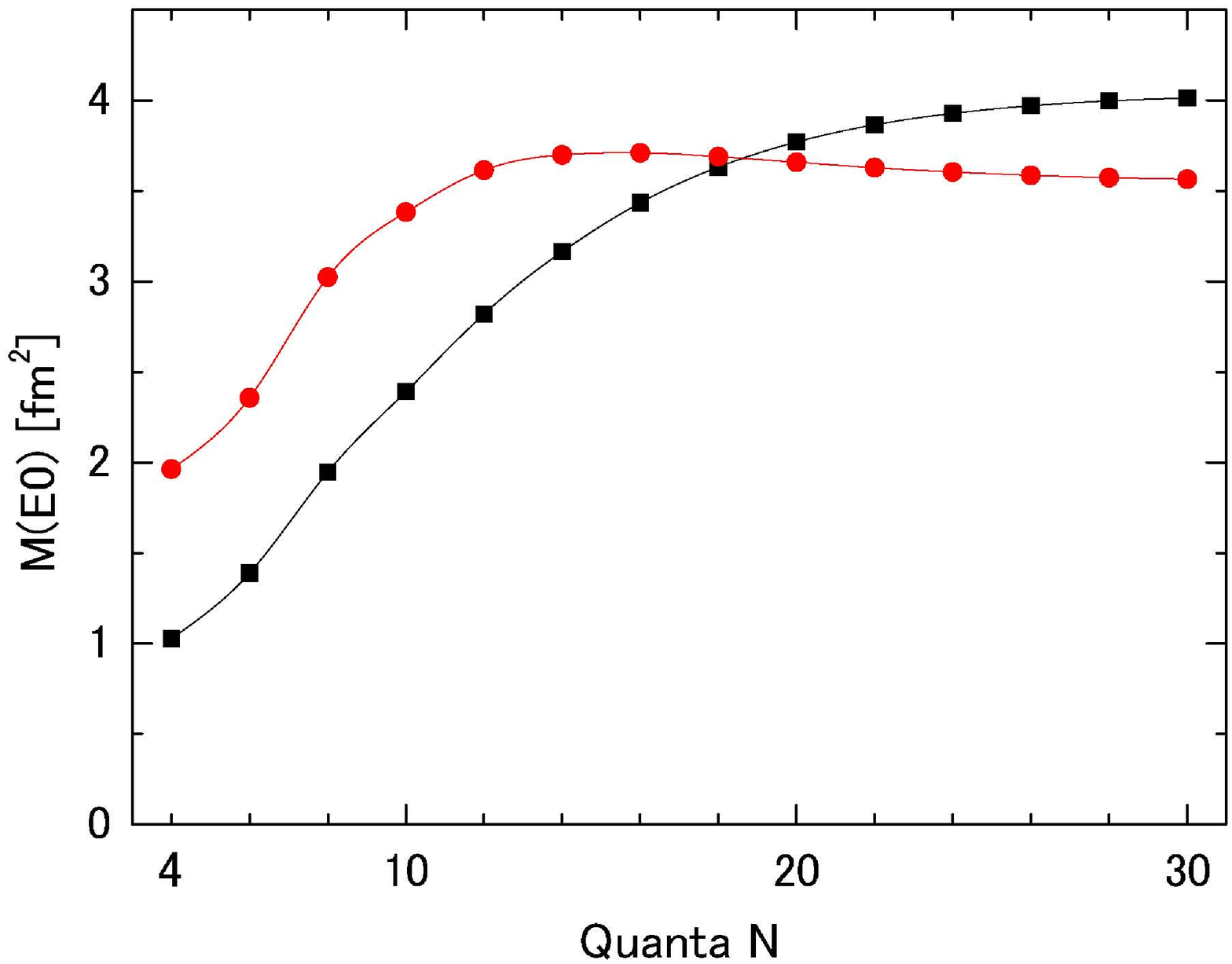}
\caption{
Dependence of the monopole strengths $M_{N}({\rm E0})$ 
 on the model space of the ground state wave function characterized as quanta $N$
 [see Eq.~(\ref{eq:monopole_N_dependence})].
The square and circle points  correspond to $M_{N}({\rm E0}; {0^+_1} \rightarrow {0^+_2})$
 and $M_{N}({\rm E0}; {0^+_1} \rightarrow {0^+_3})$, respectively.
}
\label{fig:1}
\end{center}
\end{figure}

In order to study the effect of the ground state correlation we make use 
 of the $^{12}$C + $\alpha$ OCM calculation.  
We repeat the same calculation as Ref.~\cite{suzuki}.  
Of course we adopt the same effective nuclear force.  
We express by $|0_2^+ \rangle$ and $|0_3^+ \rangle$ the obtained 
 OCM wave functions of the $0_2^+$ and $0_3^+$ states, respectively. 
Next, by $|0_1^+; N \rangle$ we express the wave function of the ground state 
 ($0_1^+$) which is calculated within the limited cluster model space 
 where the highest number of the total oscillator quanta of the basis 
 wave function is $N$ ($N$ = 4, 6, $\cdots$, 30 ).  
The wave function $|0_1^+; N=4 \rangle$ is just the closed shell wave function
 without any ground state correlation. 
As $N$ becomes larger the wave function $|0_1^+; N \rangle$ contains
 more amount of ground state correlation. 
Since $|0_2^+ \rangle$ and $|0_3^+ \rangle$ are not orthogonal to 
$|0_1^+; N \rangle$, we construct the orthogonalized wave functions 
as follows 
\begin{eqnarray}
  |\widetilde{0_k^+} \rangle = N_k (1 - |0_1^+; N \rangle 
    \langle 0_1^+; N| ) |0_k^+ \rangle, \ \ ( k=2, 3),   
\end{eqnarray}
where $N_k$ is the normalization constant.  We calculate the monopole matrix 
element between $|\widetilde{0_k^+} \rangle$ and $|0_1^+; N \rangle$ 
\begin{eqnarray}
  M_N({\rm E0}, 0_1^+ - 0_k^+) = \langle  0_1^+; N | \frac{1}{2} \sum_{i=1}^{16} 
  ( \Vec{r}_i - \Vec{r}_G )^2 |\widetilde{0_k^+} \rangle,  
  \label{eq:monopole_N_dependence}
\end{eqnarray}
and study the dependence on $N$ of this quantity.  
Figure~\ref{fig:1} shows the calculated results of $M_N({\rm E0}, 0_1^+ - 0_k^+), ( k=2, 3)$
 as a function $N$.  
The values of $M_N({\rm E0}, 0_1^+ - 0_k^+)$ for $N=4$ are the monopole matrix 
 elements without ground state correlation which we discussed in detail in the above. 
They are $M_{N=4}({\rm E0}, 0_1^+ - 0_2^+)$ = 1.39 fm$^2$ and 
 $M_{N=4}({\rm E0}, 0_1^+ - 0_3^+)$ = 2.36 fm$^2$. 
These values are very close to the values given in Eqs.~(\ref{eq:result3}) and (\ref{eq:result4})
 for which the normalization constants $N_k$ are not accounted.  
We clearly see that the values of $M_N({\rm E0}, 0_1^+ - 0_k^+)$ grow almost monotonously as $N$ 
 becomes larger. 
At $N$ = 30, the values are already converged, and they are 
$M_{N=30}({\rm E0}, 0_1^+ - 0_2^+)$ = 4.01 fm$^2$ and $M_{N=30}({\rm E0}, 0_1^+ - 0_3^+)$ = 3.56 fm$^2$.  
These converged values are very close to the observed values, and the difference
 from the observed values is only within 13~\%. 
Thus we have shown that by taking into account the ground state correlation 
 theoretical values are improved as to attain the reproduction of the data 
 within a factor of $1.13$.  

An important reason why the ground state correlation enhances the monopole strengths
 is explained as follows.
We study the deviation of the ground-state wave function $|0^{+}_{1} \rangle$ obtained
 by the $^{12}$C+$\alpha$ OCM from the doubly closed shell wave function.
For this purpose we define a modified doubly closed shell model wave function 
 $\Phi_{0^+}(\beta)$ ($\beta$ denoting the size parameter of the $^{12}$C+$\alpha$
 relative wave function)
 and calculate the squared overlap of it
 with the OCM ground state wave function
 $|0^{+}_{1} \rangle$ obtained with the full model space,
\begin{eqnarray}
&& {\Phi_{0^+}(\beta)}
    = N_{g0}(\beta) \frac{1}{\sqrt{{}_{16}C_4}} 
        {\cal A} \{ [{\cal R}_{40}(\Vec{r},\beta) 
        \phi_{L=0}({}^{12}{\rm C})]_{J=0} \phi(\alpha) \}, \label{eq:clshla_beta} \\
&& P(\beta/(3\nu_{N})) = \left| \langle  {\Phi_{0^+}(\beta)}  |  {0^+_1} \rangle \right|^{2},
    \label{eq:overlap}
\end{eqnarray}  
where $N_{g0}(\beta)$ is the normalization constant.
When $\beta$ is equal to $3\nu_{N}$, the wave function $\Phi_{0^+_1}(\beta=3\nu_{N})$
 is equivalent to the doubly closed shell model wave function in Eq.~(\ref{eq:clshla}), which
 originally has the $\alpha$ cluster degree of freedom or sort of like a seed of $\alpha$
 clustering, as discussed in Sec.~\ref{subsection:Bayman_Bohr}. 
For $\beta < 3\nu_{N}$, ${\Phi_{0^+}(\beta)}$ expresses a wave function
 in which the $^{12}$C+$\alpha$ relative motion or 
 the seed of $\alpha$ clustering is swollen in comparison with those in the doubly closed shell
 model wave function. 
Thus, the study of the dependence of the squared overlap $P$ on
 the parameter $\beta/(3\nu_{N})$ gives a rough indication on the degree of the deviation
 from the doubly closed wave function, i.e.~the degree of $\alpha$ clustering activated
 in the ground state.
We found that $P$ has the maximum value of 0.958 at $\beta/(3\nu_{N})=0.847$
 [vs.~$P=0.890$ at $\beta/(3\nu_{N})=1$].
The result of $\beta/(3\nu_{N})=0.847$ means that in the ground state wave function
 $|0^{+}_{1} \rangle$ the $^{12}$C+$\alpha$ relative motion or the seed of $\alpha$
 clustering is definitely swollen in comparison with those in the doubly closed shell model wave function.
Thus the ground state correlation makes the structure of the ground state closer
 to the $^{12}$C+$\alpha$ cluster structure in the $0^{+}_{2}$ and $0^{+}_{3}$ states,
 and then the monopole strengths become significantly larger in comparison with those 
 with no ground state correlation.

\subsection{Monopole transition matrix elements in $^{12}$C}\label{results:12C}

The analytical expression of the monopole transition matrix element 
$M({\rm E0},0^+_1-0^+_2)$ is demonstrated in Eq.~(\ref{eq:form3}),  which 
depends on the nucleon size parameter $\nu_N$ and width parameter of 
the Hoyle state $\gamma$. The expression of the monopole matrix element 
consists of three parts like the case of $^{16}$O, and the dominant 
part is the radial integral referring the relative motions among three 
$\alpha$ clusters, 
${\langle R_{40}(r,\nu_N) | r^2 | R_{60}(r,\nu_N) \rangle} 
= \sqrt{21/8}/\nu_N$. The strength of the radial integral is a few times  
larger than the single particle monopole strengths, 
$\langle 0s |r^2| 1s \rangle$ and $\langle 0p |r^2| 1p \rangle$. 

In the present study we use the value $\nu_N=0.168\ {\rm fm}^{-2}$, which 
reproduces the observed rms radius of $^{12}$C with the SU(3) shell model 
wave function in Eqs.~(\ref{eq:c12a})$\sim$(\ref{eq:c12c}). 
The monopole matrix element in Eq.~(\ref{eq:form3}) which is expressed 
as $M({\rm E0},0_2^+ - 0_1^+)=\xi_5\times 0.882/\nu_N$ ${\rm fm}^2$ contains 
$\xi_5$ which depends on $\gamma$. 
In Table~\ref{tab:2}, we display $\xi_5$ and $M({\rm E0},0_2^+ - 0_1^+)$ calculated 
at several $\gamma$ values.  According to Ref.~\cite{funaki}, we should use 
the value of $\gamma \approx 0.018$ fm$^{-2}$.  With this value of $\gamma$, 
the Bose-condensate wave function has a very large (almost 100~\%) overlap 
with the full solution of the $3\alpha$ RGM for the Hoyle state and gives 
an rms radius 3.8 fm for the Hoyle state.  
For $\gamma$ = 0.018 fm$^{-2}$, we obtain $\xi_5=0.191$, which is relatively 
small in comparison with $\eta_6=0.379$ and $\zeta_6=0.570$ in the case of 
$^{16}$O in Sec.~\ref{results:16O}.  This leads to 
\begin{eqnarray}
  M({\rm E0},0_2^+ - 0_1^+) =  0.19 \times \frac{0.882}{\nu_N} = 1.3 {\rm fm}^2, 
\end{eqnarray}
which is of the same order of magnitude as the observed 
value ($5.4 \pm 0.2$ ${\rm fm}^2$~\cite{ajzen86}) but reproduces only about 
25 \% in comparison with that. This value of about 25 \% is a little bit 
smaller in contrast to that of the $^{16}$O case (see the previous subsection 
\S\ref{results:16O}) in which our simple estimates are larger than about 
40\% of the experimental data.
 
We should note that in more realistic situation the description of the ground 
state adopted here for $^{12}$C using the SU(3) shell model is not necessarily 
good and a deviation from the SU(3) shell model representation should be 
taken into account~\cite{supple}.  
According to the structure study of $^{12}$C with the $3\alpha$ orthogonality
 condition model (OCM)~\cite{yamada}, the SU(3)$(\lambda,\mu)=(0,4)$ component
 with the lowest oscillator quantum ($N_{TOT}=8$) is only about
 60~\% in the ground state. 
The smallness of the SU(3)$(\lambda,\mu)=(0,4)$ component with the lowest 
oscillator quantum ($N_{TOT}=8$) is in contrast to the $^{16}$O case, in 
which the ground state is well described by the doubly closed shell model 
wave function $(0s)^4(0p)^{12}$:~The SU(3)$(\lambda,\mu)=(0,0)$ component 
with the lowest quantum ($N_{TOT}=12$) of the $^{16}$O ground state is as 
large as about 90\%~\cite{suzuki}. 

Here we demonstrate the effect of the ground state correlation to the 
monopole matrix element by adopting the following wave function for the 
ground state~\cite{funaki}:
\begin{equation}
\Psi_G({\widetilde \gamma},\nu_N)={\cal N}_G \sqrt{\frac{4!4!4!}{12!}}
{\cal A}[\exp\{-{\widetilde \gamma}(2\Vec{s}^2+\frac{8}{3}\Vec{t}^2) \} 
\phi(\alpha_1)\phi(\alpha_2)\phi(\alpha_3)],\label{eq:Psi_G}
\end{equation}
where ${\cal N}_G$ is normalization constant.  This wave function is 
called THSR wave function and depends on two 
parameters ${\widetilde \gamma}$ and $\nu_N$.  We impose the condition  
that this wave function reproduces the observed rms radius, 2.47 fm, of 
the ground state.  Then, the ratio ${\widetilde \gamma}/\nu_N$ is the only  
parameter which describes the property of the ground state.  
It is noted that $\Psi_G({\widetilde \gamma},\nu_N)$ with ${\widetilde \gamma}/\nu_N=1$ 
 agrees with the SU(3) $(\lambda,\mu)=(0,4)$ shell model wave function
 in Eqs.~(\ref{eq:c12a})$\sim$(\ref{eq:c12c}), demonstrating directly
 that the wave function originally has sort of like a seed of $3\alpha$
 clustering. 
Taking the ${\widetilde \gamma}$ value a little smaller than $\nu_N$, 
 $\Psi_G({\widetilde \gamma},\nu_N)$ deviates from the SU(3) shell model 
 wave function and the slightly more relaxed spatial clustering of 
 $3\alpha$ clusters than the SU(3) wave function is induced in the ground 
 state, i.e.~$\Psi_G({\widetilde \gamma},\nu_N)$ expresses sort of like
 a generalized SU(3) wave function in which the seed of $3\alpha$
 clustering are slightly swollen in comparison with the original SU(3) wave function. 
This is the ground state correlation taking into account here,
 which is similar to the case in $^{16}$O as discussed in Sec.~\ref{results:16O}. 
In Ref.~\cite{funaki}, it is reported that the ground 
 state wave function of the $3\alpha$ RGM calculation of 
 Refs.~\cite{kami} and \cite{uega} can be well approximated by this kind of 
 wave function.  
The amount of the $3\alpha$-like ground state correlation, 
 thus, can be characterized by the ratio ${\widetilde \gamma}/\nu_N$, 
 which should be less than or equal to unity.  
In the $3\alpha$ cluster model~\cite{supple,funaki,kami,uega,yamada}, 
 the nucleon size parameter $\nu_N$ is usually chosen to reproduce
 the rms radius of $\alpha$ cluster, $\nu_N=0.275~{\rm fm}^{-2}$ which
 is larger than that for the SU(3) shell 
 model wave function ($\nu_N=0.168~{\rm fm}^{-2}$) shown above. 
The estimation of ${\widetilde \gamma}/\nu_N$ for the ground state wave 
 function given by Ref.~\cite{funaki} is as small as 
 ${\widetilde \gamma}/\nu_N\sim0.29$.
This small value indicates that 
 the ground state of $^{12}$C has a significant amount of the $3\alpha$ 
 correlation. 
Below we change the value of the parameter 
 ${\widetilde \gamma}/\nu_N$ from $1.0$ down to $0.27$.  

The wave function of the Hoyle state is constructed so as to be orthogonal 
to the ground state wave function $\Psi_G$ in Eq.~(\ref{eq:Psi_G}) and 
is given as follows:
\begin{eqnarray}
&&\Psi_H(\gamma,{\widetilde \gamma},\nu_N)={\cal N}_H (1-P) 
 \sqrt{\frac{4!4!4!}{12!}} {\cal A} [\exp \{ -\gamma (2\Vec{s}^2 + 
 \frac{8}{3}\Vec{t}^2) \}\phi(\alpha_1)\phi(\alpha_2)\phi(\alpha_3)], 
 \label{eq:phi_H_modify}\\
&& P \equiv | \Psi_G({\widetilde \gamma},\nu_N) \rangle \langle 
  \Psi_G({\widetilde \gamma},\nu_N) |, 
\end{eqnarray}
where ${\cal N}_H$ is normalization constant. 
The width parameter $\gamma$ is determined so as to reproduce the rms 
radius of the Hoyle state, 3.8 fm.  The Hoyle state of $^{12}$C ($0^+_2$) 
is known to have a dilute $3\alpha$ condensate structure with the nuclear 
radius of about 4 fm.  
This exotic structure of the Hoyle state was found to be described 
simply~\cite{funaki} with a single $\alpha$-condensate wave function 
given in Eq.~(\ref{eq:phi_H_modify}).  The monopole matrix element given as 
\begin{eqnarray}
{M({\rm E0},0^+_1-0^+_2})=
 {\langle \Psi_G({\widetilde \gamma},\nu_N) | \frac{1}{2}\sum_{i=1}^{12} 
  (\Vec{r}_i-\Vec{r}_G)^2 | \Psi_H(\gamma,{\widetilde \gamma},\nu_N) 
  \rangle},
 \label{eq:E0_HG}
\end{eqnarray} 
depending only on the parameter ${\widetilde \gamma}/\nu_N$. 
 
Table~\ref{tab:3} shows the values of the monopole matrix elements 
[Eq.~(\ref{eq:E0_HG})]  calculated at several ${\widetilde \gamma}/\nu_N$ 
values.  We see that the monopole matrix element increases as the ratio 
${\widetilde \gamma}/\nu_N$ decreases from unity, namely as the 
$3\alpha$-like correlation becomes stronger in the ground state. 
This can be reasonably understood from the fact that the ground state wave 
function $\Psi_G$ with stronger $3\alpha$-like correlation has larger 
$3\alpha$-cluster component which makes larger the overlap with the Hoyle 
state wave function $\Psi_H$ with the dilute $3\alpha$ cluster structure, 
and then the monopole matrix element becomes larger. 
At the value of ${\widetilde \gamma}/\nu_N\sim0.27$, the monopole matrix 
element is about $4.0$ fm$^2$, which is about three times larger than that 
for ${\widetilde \gamma}/\nu_N=1$, and is closer to the observed value 
$5.4\pm0.2$ fm$^2$. It is noted that ${\widetilde \gamma}/\nu_N\sim0.27$ 
gives the nucleon size parameter $\nu_N\sim0.26~{\rm fm}^{-2}$  which 
corresponds to the value used usually in the microscopic $3\alpha$ 
cluster model calculations~\cite{supple,funaki,kami,uega,yamada}.  
Without the ground state correlation the calculated monopole 
value is smaller than the observed value by a factor of 4.15 but 
now with inclusion of the ground state correlation the calculated monopole 
value changed to be smaller only by a factor of 1.35 than the observed 
value.

\section{Discussions and Summary}\label{summary}

The monopole transitions from cluster states to ground states in light 
 nuclei are rather large which is comparable with the single particle strength. 
The single particle estimate of the monopole transition is based on 
 the assumption that the excited state has a one-particle one-hole excitation 
 from the ground state.  
However, the cluster structure is very different 
 from the shell-model-like structure of the ground state, and its state is 
 described as a superposition of many-particle many-hole configurations when 
 it is expanded by shell model configurations.   
This means that in the excited state with a cluster structure, 
 the component of a one-particle one-hole excitation from the ground state
 configuration is expected to be very small.   
Therefore the observation of rather large monopole strengths 
 for cluster states which are comparable with single particle strength 
 looks {\it not} to be easy to explain.  
Under this kind of understanding it has been often regarded that the monopole
 transition occurs through the mixing of shell model wave function
 $|\rm{shell} \rangle$ and the cluster model wave function $|\rm{cluster} \rangle$ 
\begin{eqnarray}
 && |\rm{Ground} \rangle = \alpha |\rm{shell} \rangle + \beta 
    |\rm{cluster} \rangle, \\ 
 && |\rm{Excited} \rangle = -\beta |\rm{shell} \rangle + \alpha 
    |\rm{cluster} \rangle.  
\end{eqnarray}
Since it is assumed that the monopole operator $O_M$ does not connect 
 $|\rm{cluster} \rangle$ and $|\rm{shell} \rangle$, 
 $\langle \rm{cluster} | {\it O_M} | \rm{shell} \rangle$ = 0, 
 the monopole matrix element is considered to come
 from the diagonal matrix elements (for example see Ref.~\cite{bertsch}), 
\begin{eqnarray}
  \langle \rm{Excited} | {\it O_M} |\rm{Ground} \rangle = \alpha \beta ( 
   \langle \rm{cluster} | {\it O_M} |\rm{cluster} \rangle - \langle 
   \rm{shell} | {\it O_M} |\rm{shell} \rangle ).  
\end{eqnarray}
Our explanation of the strong monopole transition between ground state and 
excited cluster states is quite different from this explanation.  We 
insist that the order of magnitude of the strong monopole transition is 
given by the matrix element $\langle \rm{cluster} | {\it O_M} | \rm{shell} 
\rangle \ne$ 0, 
\begin{eqnarray}
  \langle \rm{Excited} | {\it O_M} |\rm{Ground} \rangle \approx 
   \langle \rm{cluster} | {\it O_M} |\rm{shell} \rangle, \  ({\rm for} 
   \ {\rm order} \ {\rm of} \ {\rm magnitude}).
\end{eqnarray} 
Our argument is based on the Bayman-Bohr theorem which says that the SU(3) 
 shell-model wave function which describes rather well the structure of the 
 ground state of light nuclei is equivalent in most cases to cluster model wave 
 function. 
The implication of this theorem is that the clustering degree of 
 freedom is already embedded even in the shell model wave function. 
In the present study the monopole excitation of the ground state to cluster 
 states is understood as just the excitation of the inter-cluster relative motion in the 
 ground state to the inter-cluster relative motion in excited cluster states. 
This resembles the monopole excitation of the single nucleon motion. 
Our understanding was explicitly shown to be true by deriving the analytical 
 expressions of the monopole matrix elements.     

In this paper we analyzed the monopole transitions in $^{16}$O between the 
ground state and $^{12}$C + $\alpha$ cluster states ($0^+_2$ and $0^+_3$) 
together with the one in $^{12}$C between the ground state and $3\alpha$ 
cluster state ($0^+_2$: Hoyle state).  According to the Bayman-Bohr theorem, 
the doubly closed shell model wave function of $^{16}$O, which has the 
SU(3)($\lambda \mu$)=(00) symmetry and total quanta $N_{\rm TOT}=12$, is 
equivalent to the $^{12}$C ($0^+$) + $\alpha$ cluster wave function as 
well as $^{12}$C ($2^+$) + $\alpha$ with orbital angular momentum of 
the inter-cluster relative motion $L_r=0$ and $2$, respectively.  
The number of oscillator quanta of the inter-cluster relative motion 
$N_r$ is 4.  Similarly, the ground state wave function of $^{12}$C  with 
SU(3) ($\lambda\mu$) = (04) and $N_{\rm TOT}=8$ is equivalent to the 
$3\alpha$ cluster wave function.  The number of oscillator quanta of the 
inter-cluster relative motion with respect to each of two Jacobi 
coordinates is $N_r$ = 4.  On the other hand, the $0^+_2$ [$0^+_3$] 
state of $^{16}$O has a $^{12}$C($0^+$)+$\alpha$ cluster structure 
[$^{12}$C($2^+$)+$\alpha$]  with the relative orbital angular momentum 
$L_r=0$ [$L_r=2$].  The $0^+_2$ state of $^{12}$C has a $3\alpha$ cluster 
structure with  $S$-wave relative angular momenta referring to two 
Jacobi coordinates for $3\alpha$ clusters. 

The analytical expressions of the monopole matrix elements we derived 
for the above transitions in $^{16}$O and $^{12}$C are composed of three 
factors.  The most important factor is the radial integrals with harmonic 
oscillator wave functions $\langle N_r=6,L_r | r^2 | N_r=4,L_r \rangle$ 
with $L_r=0$ or 2. The values of these integrals are a few times larger 
than the single particle monopole transition matrix elements in {\it p}-shell 
nuclei, ${\langle 1s | r^2 | 0s \rangle}$ = 
${\langle N_r=2,L_r=0 | r^2 | N_r=0, L_r=0 \rangle}$ and 
${\langle 1p | r^2 | 0p \rangle}$ = 
${\langle N_r=3,L_r=1 | r^2 | N_r=1,L_r=1\rangle}$. 
The second factor is 
the amplitude (not squared amplitude) of the $2\hbar \omega$ - excited 
harmonic oscillator wave function in the cluster states which is denoted as 
$\eta_6$ for $M({\rm E0},0^+_1-0^+_2)$ and $\zeta_6$ for $M({\rm E0},0^+_1-0^+_3)$ 
in $^{16}$O and $\xi_5$ for $M({\rm E0},0^+_1-0^+_2)$ in $^{12}$C.  They are 
not so small; $\eta_6$ = 0.38, $\zeta_6$ = 0.57, and $\xi_5$ = 0.19. 
The third factor is due to the antisymmetrization among nucleons, which 
is denoted as $\sqrt{\tau_{0,4}/\tau_{0,6}}$ or 
$\sqrt{\tau_{2,4}/\tau_{2,6}}$ in $^{16}$O and 
$\sqrt{\langle F_4 \rangle / \langle F_5 \rangle}$ in 
$^{12}$C.  Since the quantities with strong antisymmetrization effect 
are contained in the form of ratio, the third factor has magnitude close 
to unity.  As is expected from the analytical expressions, the calculated 
numerical values of the monopole matrix elements were shown to have the same 
order of magnitude as the observed values which are comparable with the 
single nucleon strength. 

Although the calculated values of the monopole matrix elements without 
 ground state correlation surely reproduce the order of magnitude of the 
 observed values, when compared with the data in detail, they are a few times 
 smaller than the observed values. In the case of $^{16}$O, two kinds of 
 theoretical values of $M({\rm E0},0^+_1-0^+_2)$ are 60 \% and 38 \% of the 
 observed values, respectively,
 while those of $M({\rm E0},0^+_1-0^+_3)$ are 96 \% and 63 \%. 
In the case of $^{12}$C, theoretical value of $M({\rm E0},0^+_1-0^+_2)$ is 25 \% 
of the observed value.  Therefore we next investigated the effect of the 
ground state correlation to the monopole matrix elements. 
The ground state correlation we considered was the one caused by the 
activation of the clustering degree of freedom described by Bayman-Bohr 
theorem. In the calculation of the monopole strength without ground state 
correlation, we only considered the clustering degree of freedom rather 
in a static way.  Namely we did not consider the dynamical effect of the 
clustering degree of freedom which excites the ground state configuration 
toward including higher quantum configurations.  We know that the 
clustering degree of freedom described by Bayman-Bohr theorem has the 
physical reality because we observe many excited cluster states which are 
formed by exciting the clustering degree of freedom embedded in the ground 
state. Therefore taking into account the ground state correlation caused 
by the clustering degree of freedom described by Bayman-Bohr theorem is 
very natural and should be studied.

The investigation of the effect of the ground state correlation to the 
 monopole strength in $^{16}$O was made in the framework of the $^{12}$C + $\alpha$ OCM. 
It is because, in discussing the monopole strength without ground state correlation
 in Sec.~\ref{formulation}, we used the results of the $^{12}$C + $\alpha$ OCM
 in Ref.~\cite{suzuki}. 
We repeated the same calculation as one in Ref.~\cite{suzuki}. 
We found that 1)~increasing the amount of the ground state correlation, 
 the monopole strengths are growing almost monotonously, and 
 2)~at a full amount of the ground state correlation, the monopole strengths
 are reproduced within a factor of $1.13$ in comparison with the observed values.  
The reason why the ground state correlation enhances the monopole strengths
 was discussed with a simple approach. 
In the case of $^{12}$C, the investigation of the effect of the ground state
 correlation to the monopole strength was performed by expressing the ground state
 with the so-called THSR wave function~\cite{tohsaki}. 
This wave function has two parameters ${\widetilde \gamma}$ and $\nu_N$.  
When ${\widetilde \gamma}$ = $\nu_N$, 
 the wave function is just equal to the SU(3) wave function with 
 $N_{TOT}=8$ and $(\lambda, \mu) = (0, 4)$.  
As we make the ratio ${\widetilde \gamma} / \nu_N$ smaller than unity, the wave 
 function contains more amount of the ground state correlation. 
We found that at a full amount of the ground state correlation, 
 the monopole strength is reproduced within a factor of $1.35$ in comparison 
 with the observed value. 

The implication of the Bayman-Bohr theorem has been misunderstood such that 
 the cluster model description is rather unnecessary, since a cluster model 
 wave function is equivalent to a shell model wave function.  
The existence of cluster states especially as excited states is well 
 established these days. 
Thus, the implication of the Bayman-Bohr theorem should be understood
 straightforwardly as follows.  
If the ground state is well described by an SU(3) shell model wave 
 function equivalent to a cluster model wave function, the ground state 
 possesses two different characters simultaneously, shell-model-state 
 character and cluster-model-state character.  
This means that the ground state has mean-field degree of freedom
 and clustering degree of freedom simultaneously.
Both of them can be excited, when the nucleus is stimulated by an external field.  
The monopole excitation to excited cluster states demonstrates us directly
 the evidence that the clustering degree of freedom is embedded in the ground state.
In this paper we showed that the clustering degree of freedom embedded
 in the ground state can reproduce the order of magnitude of the monopole strength
 even without taking into account the ground state correlation. 
Moreover it was demonstrated that, if we take into account the 
 ground state correlation activating the clustering degree of freedom
 described by the Bayman-Bohr theorem, 
 the monopole strengths are reproduced within a factor of $1.13$ in $^{16}$O
 and within a factor of $1.35$ in $^{12}$C, in comparison with the observed values.   

Our present study ascertains that the monopole transition between cluster 
 and ground states in light nuclei is generally strong as to be comparable 
 to the  single particle strength. 
The measurement of strong monopole transitions or excitations, 
 therefore, is in general very useful for the study of cluster states. 

One of the authors (Y.~F.) is grateful for the financial assistance 
 from the Special Postdoctoral Researcher Program of RIKEN.

\appendix

\section{The energy-weighted sum rule of monopole transition by the use of 
the Jacobi coordinate} \label{sumrule}

We discuss here the energy-weighted sum rule of the monopole transition.  
The sum rule is written as follows 
\begin{eqnarray}
  && \sum_k |\langle k | \frac{1}{2} \sum_{i=1}^A
    (\Vec{r}_i - \Vec{r}_G )^2 | g \rangle|^2 ( E_k - E_g ) = 
    \frac{\hbar^2}{2m} A R^2_{\rm rms},  \\
  && R^2_{\rm rms} = \frac{1}{A} \langle g | \sum_{i=1}^A
    (\Vec{r}_i - \Vec{r}_G )^2 | g \rangle,
\end{eqnarray}
where $| g \rangle$ and $E_g$ stand for the ground state and 
 its energy, respectively, $| k \rangle$ and $E_k$ represent 
 the $k$-th excited state and its energy, respectively, and 
 $\Vec{r}_G$ stands for the center-of-mass coordinate. 
 
In $^{16}$O, the observed value of $R_{\rm rms}$ is 2.67 fm and then 
 the energy-weighted sum rule value $(\hbar^2/2m) 16 R^2_{\rm rms}$ 
 is 2361 fm$^4 \cdot$MeV.  
In the case of the $0^+_2$ state at 6.05 MeV which has 
 $M({\rm E0},0^+_1-0^+_2)_{\rm exp}$ = 3.55 fm$^2$, the energy-weighted monopole 
 transition strength is $(3.55)^2 \times 6.05 = 76.3$ fm$^4 \cdot$MeV. 
This value is 3.2 \% of the energy-weighted sum rule value. 
In the case of the $0^+_3$ state at 12.05 MeV which has 
 $M({\rm E0},0^+_1-0^+_3)_{\rm exp}$ = 4.03 fm$^2$, the energy-weighted 
 monopole transition strength is $(4.03)^2 \times 12.05 = 196$ 
 fm$^4 \cdot$MeV. This value is 8.3 \% of the energy-weighted 
 sum rule value.  
The sum of the energy-weighted monopole transition strengths 
 of $0^+_2$ and $0^+_3$ states is 11.5 \% of the energy-weighted 
 sum rule value. 
In $^{12}$C, the observed value of $R_{\rm rms}$ is 2.37 fm and then 
 the energy-weighted sum rule value $(\hbar^2/2m) 12 R^2_{\rm rms}$ 
 is 1395 fm$^4 \cdot$MeV.  
In the case of the $0^+_2$ state at 7.66 MeV which has 
 $M({\rm E0},0^+_1-0^+_2)_{\rm exp}$ = 5.4 fm$^2$, 
 the energy-weighted monopole transition strength is 
 $(5.4)^2 \times 7.66 = 223$ fm$^4 \cdot$MeV. 
This value is 16 \% of the energy-weighted sum rule value. 
These percentage values show that the strength of the monopole 
 transition or excitation to cluster states shares an appreciable 
 portion of the energy-weighted sum rule value. 

The formula of the energy-weighted sum rule of monopole transition is 
obtained by calculating the double commutator of the monopole transition 
operator $O_M$ and the system Hamiltonian $H$
\begin{equation}
  [ O_M, [ H, O_M ]],  \ \ \ O_M = \frac{1}{2} \sum_{i=1}^A
    (\Vec{r}_i - \Vec{r}_G )^2.
\end{equation}
The calculation of the double commutator looks tedious due to 
the existence of the center-of-mass coordinate $\Vec{r}_G$ but 
it can be made very easily by using the Jacobi coordinate.  
For the Hamiltonian $H$ with momentum-independent interaction, $H$ can be 
replaced by the kinetic energy operator $K$
\begin{equation}
  [ O_M, [ H, O_M ]] = [ O_M, [ K, O_M ]],  \ \ \  K = \frac{-\hbar^2}{2m} 
  \sum_{i=1}^A \left( \frac{\partial}{\partial \Vec{r}_i} \right)^2 
  - \frac{-\hbar^2}{2Am} \left( \frac{\partial}{\partial \Vec{r}_G} \right)^2. 
\end{equation}

Now we introduce the normalized Jacobi coordinates as 
\begin{eqnarray}
  \Vec{x}_j &=&  \sqrt{\frac{j}{j+1}} \left( \frac{1}{j} \sum_{i=1}^j 
      \Vec{r}_i - \Vec{r}_{j+1} \right), \ \ \ j = 1 \sim A-1, \\ 
  \Vec{x}_A &=& \sqrt{\frac{1}{A}} \sum_{i=1}^A \Vec{r}_i = 
      \sqrt{A} \Vec{r}_G.
\end{eqnarray}
One can easily check that the linear transformation from 
$\{ \Vec{r}_i, i=1 \sim A \}$ to $\{ \Vec{x}_j, j=1 \sim A \}$ is unitary. 
Therefore we have 
\begin{eqnarray}
  && \sum_{i=1}^A \Vec{r}_i^2 = \sum_{j=1}^A \Vec{x}_j^2,   \\ 
  && \sum_{i=1}^A (\Vec{r}_i - \Vec{r}_G)^2 = \sum_{i=1}^A \Vec{r}_i^2 - 
    A \Vec{r}_G^2 = \sum_{j=1}^{A-1} \Vec{x}_j^2,   \\ 
  && \sum_{i=1}^A \left( \frac{\partial}{\partial \Vec{r}_i} \right)^2 
    = \sum_{j=1}^A \left( \frac{\partial}{\partial \Vec{x}_j} \right)^2, \\ 
  &&  \sum_{i=1}^A \left( \frac{\partial}{\partial \Vec{r}_i} \right)^2 
      - \frac{1}{A} \left( \frac{\partial}{\partial \Vec{r}_G} \right)^2
    = \sum_{j=1}^{A-1} \left( \frac{\partial}{\partial \Vec{x}_j} \right)^2, 
\end{eqnarray}
Thus we have 
\begin{eqnarray}
  O_M &=& \frac{1}{2} \sum_{j=1}^{A-1} \Vec{x}_j^2,   \\
  K &=& \frac{-\hbar^2}{2m} \sum_{j=1}^{A-1} \left( \frac{\partial}
     {\partial \Vec{x}_j} \right)^2, 
\end{eqnarray}

When we use the above expressions of $O_M$ and $K$ by the normalized Jacobi 
coordinates, we can easily obtain the following result  
\begin{equation}
  [ O_M, [ H, O_M ]] = [ O_M, [ K, O_M ]] = \frac{\hbar^2}{m} 
    \sum_{j=1}^{A-1} \Vec{x}_j^2 = \frac{\hbar^2}{m} \sum_{i=1}^A 
    (\Vec{r}_i - \Vec{r}_G)^2.
\end{equation}
We thus have the formula of the energy-weighted sum rule of monopole 
transition as follows 
\begin{eqnarray}
  && \frac{\hbar^2}{2m} \langle g | \sum_{i=1}^A (\Vec{r}_i - \Vec{r}_G)^2 | g 
     \rangle = \frac{1}{2} \langle g | [ O_M, [ H, O_M ]] | g \rangle  \\ 
  && \ \ = \sum_k |\langle k | \frac{1}{2} \sum_{i=1}^A
    (\Vec{r}_i - \Vec{r}_G )^2 | g \rangle|^2 ( E_k - E_g ). 
\end{eqnarray}

\section{Description of $^{16}$O and $^{12}$C ground states with SU(3) wave function} \label{Baymann_Bohr} 

The total number of the oscillator quanta $N_{TOT}$ possessed by the 
 doubly closed shell wave function of $^{16}$O is $N_{TOT}=12$.
For $N_{TOT}=12$, it is possible to construct many $^{12}$C+$\alpha$ 
 cluster wave functions with various SU(3) symmetry $(\lambda,\mu)$,
 ${\cal A} \{ [{\cal R}_4(\Vec{r}) \phi({}^{12}{\rm C})]_{(\lambda,\mu)} \phi (\alpha) \}$.  
These wave functions, however, become all zero except for $(\lambda,\mu)=(0,0)$,
 because of the nature of the doubly closed wave function. 
Using the following relation
\begin{equation}
  [{\cal R}_{4L}(\Vec{r}) \phi_L({}^{12}{\rm C})]_{J=0} = \sum_{(\lambda,\mu)} 
  \langle (4,0)L, (0,4)L ||(\lambda,\mu)0 \rangle 
  [{\cal R}_4(\Vec{r}) \phi({}^{12}{\rm C})]_{(\lambda,\mu)}, 
\end{equation}
we have 
\begin{eqnarray}
  && {\cal A} \{ [{\cal R}_{4L}(\Vec{r}) \phi_L({}^{12}{\rm C})]_{J=0} 
    \phi (\alpha) \} \nonumber \\ 
  && \hspace{1.5cm} = \sum_{(\lambda,\mu)} 
    \langle (4,0)L, (0,4)L ||(\lambda,\mu)0 \rangle 
    {\cal A} \{ [{\cal R}_4(\Vec{r}) \phi({}^{12}{\rm C})]_{(\lambda,\mu)} 
    \phi (\alpha) \} \nonumber \\ 
  && \hspace{1.5cm} =  \langle (4,0)L, (0,4)L ||(0,0)0 \rangle 
    {\cal A} \{ [{\cal R}_4(\Vec{r}) \phi({}^{12}{\rm C})]_{(\lambda,\mu)=(0,0)} 
    \phi (\alpha) \},
\end{eqnarray}
for $L=0$, 2, and 4.
This relation is an explanation of the equalities of Eqs.~(\ref{eq:clshla})$\sim$(\ref{eq:clshlc}). 

Similar argument holds for the ground state of $^{12}$C.  
Although there can be constructed many 3$\alpha$ cluster wave functions 
 with various SU(3) symmetry $(\lambda,\mu)$, 
 ${\cal A} \{ [{\cal R}_{N_1}(\Vec{s},2\nu_N) 
 {\cal R}_{N_2}(\Vec{t},(8/3)\nu_N)]_{(\lambda,\mu)} \phi(\alpha_1) \phi(\alpha_2) 
 \phi(\alpha_3) \}$ for $N_{TOT} = N_1 + N_2 =8$ which is the lowest 
 number of the total oscillator quanta for $^{12}$C, only one wave 
 function with $(\lambda,\mu)=(0,4)$ is non-vanishing which is possible 
 for $N_1 = N_2 = 4$~\cite{hori,kato}.  
Therefore we have the following relations
\begin{eqnarray}
  && {\cal A} \{ [{\cal R}_{N_1 L}(\Vec{s},2\nu_N) {\cal R}_{N_2 L}(\Vec{t},(8/3)\nu_N)
         ]_{J=0} \phi(\alpha_1) \phi(\alpha_2) \phi(\alpha_3) \}  
        \nonumber  \\
  && \hspace{0.4cm} = \delta_{N_1,4} \delta_{N_2,4} 
     \langle (4,0)L, (4,0)L ||(0,4)0 \rangle  \nonumber  \\ 
  && \hspace{0.4cm} \times {\cal A} \{ [ {\cal R}_{N_1}(\Vec{s},2\nu_N) 
     {\cal R}_{N_2}(\Vec{t},(8/3)\nu_N)]_{(\lambda,\mu)=(0,4)} 
     \phi(\alpha_1) \phi(\alpha_2) \phi(\alpha_3) \},
\end{eqnarray}
for $L=0$, 2, and 4.
This relation is an explanation of the equalities of  
 Eqs.~(\ref{eq:c12a})$\sim$(\ref{eq:c12c}).

\section{Dependence of $M({\rm E0},0^+_1-0^+_2)$ on the width parameter $\gamma$ of 
the 3$\alpha$ condensed wave function} \label{Hoyle}

First we note 
\begin{eqnarray}
 && {\cal A} \{ {\widehat \chi}_H(\Vec{s},\Vec{t}) \phi(\alpha_1) 
   \phi(\alpha_2) \phi(\alpha_3) \} = {\cal A} 
   \{ {\widehat \chi}_{HG}(\Vec{s},\Vec{t}) \phi(\alpha_1) \phi(\alpha_2) 
   \phi(\alpha_3) \}   \nonumber \\ 
 && \hspace{1cm} - \langle  {\cal R}^{8,J=0}_{4,4,L=0}(\Vec{s},\Vec{t}) 
   | {\widehat \chi}_{HG}(\Vec{s},\Vec{t}) \rangle  
   {\cal A} \{ {\cal R}^{8,J=0}_{4,4,L=0}(\Vec{s},\Vec{t})  
   \phi(\alpha_1) \phi(\alpha_2) \phi(\alpha_3) \}.
\end{eqnarray}
Then we obtain 
\begin{equation}
  \langle {\widehat \chi}_H \rangle 
     = \langle {\widehat \chi}_{HG} \rangle - 
     \left( \langle  {\cal R}^{8,J=0}_{4,4,L=0}(\Vec{s},\Vec{t}) | 
     {\widehat \chi}_{HG}(\Vec{s},\Vec{t}) \rangle \right)^2 
     \langle  {\cal R}^{8,J=0}_{4,4,L=0} \rangle, 
\end{equation}
with the notation 
\begin{equation}
  \langle G \rangle = \langle G(\Vec{s},\Vec{t}) 
     \phi(\alpha_1) \phi(\alpha_2) \phi(\alpha_3) | {\cal A} \{ 
     G(\Vec{s},\Vec{t}) \phi(\alpha_1) \phi(\alpha_2) \phi(\alpha_3) 
     \} \rangle. \label{eq:appC_G} 
\end{equation}
Next we note 
\begin{eqnarray}
  && {\cal A} \{ {\widehat \chi}_{HG}(\Vec{s},\Vec{t}) \phi(\alpha_1) 
     \phi(\alpha_2) \phi(\alpha_3) \} = \sum_{n \ge 4}^\infty \sum_{n_1+n_2=n} 
     \langle  {\cal R}^{2n,J=0}_{2n_1,2n_2,0}(\Vec{s},\Vec{t}) 
     | {\widehat \chi}_{HG}(\Vec{s},\Vec{t}) \rangle  \nonumber \\
   && \hspace{4cm} \times {\cal A} \{ {\cal R}^{2n,J=0}_{2n_1,2n_2,0}
     (\Vec{s},\Vec{t}) \phi(\alpha_1) \phi(\alpha_2) \phi(\alpha_3) \} \\ 
   && \hspace{0.5cm} = \left( \frac{2\sqrt{\nu_N\gamma}}
    {\nu_N+\gamma} \right)^3 \sum_{n=4}^\infty \left( \frac{\nu_N-\gamma}
    {\nu_N+\gamma} \right)^n {\cal A} \{ F_n(\Vec{s},\Vec{t}) 
     \phi(\alpha_1) \phi(\alpha_2) \phi(\alpha_3) \},\label{eq:appC_HG}\\ 
  &&  F_n(\Vec{s},\Vec{t}) = \sum_{n_1+n_2=n} \sqrt{\frac{(2n_1+1)!!(2n_2+1)!!}
    {(2n_1)!!(2n_2)!!}} {\cal R}^{2n,J=0}_{2n_1,2n_2,0}(\Vec{s},\Vec{t}),   
\end{eqnarray}
where the following formula is used
\begin{equation}
  \langle R_{2n,0}(r,\beta)| \sqrt{4\pi}
  \left(\frac{2\beta'}{\pi}\right)^{\frac{3}{4}} e^{-\beta' r^2} 
  \rangle  = \sqrt{\frac{(2n+1)!!}{(2n)!!}}
  \left( \frac{2\sqrt{\beta\beta'}}{\beta+\beta'} \right)^{\frac{3}{2}} 
  \left( \frac{\beta-\beta'}{\beta+\beta'} \right)^n,  
\end{equation}  
From Eqs.~(\ref{eq:appC_G}) and (\ref{eq:appC_HG}) we obtain 
\begin{equation}
  \langle {\widehat \chi}_{HG} \rangle = 
    \left( \frac{2\sqrt{\nu_N\gamma}}{\nu_N+\gamma} \right)^6 
    \sum_{n=4}^\infty 
    \left( \frac{\nu_N-\gamma}{\nu_N+\gamma} \right)^{2n} \langle F_n \rangle. 
\end{equation}
By noticing 
\begin{eqnarray}
  && \langle  {\cal R}^{8,J=0}_{4,4,0}(\Vec{s},\Vec{t}) | 
    {\widehat \chi}_{HG}(\Vec{s},
     \Vec{t}) \rangle = \frac{5!!}{4!!} \left( \frac{2\sqrt{\nu_N\gamma}}
     {\nu_N+\gamma} \right)^3 \left( \frac{\nu_N-\gamma}{\nu_N+\gamma} 
     \right)^4, 
    \\ 
 && \langle F_4 \rangle = \left( \frac{5!!}{4!!} \right)^2 
     \langle  {\cal R}^{8,J=0}_{4,4,0} \rangle 
    \equiv \left( \frac{5!!}{4!!} \right)^2\frac{1}{(\widehat{N}_{g0})^2},
\end{eqnarray}
we obtain 
\begin{eqnarray}
  &&\langle {\widehat \chi}_H \rangle = 
    \left( \frac{2\sqrt{\nu_N\gamma}}{\nu_N+\gamma} \right)^6 
    \sum_{n=5}^\infty 
    \left( \frac{\nu_N-\gamma}{\nu_N+\gamma} \right)^{2n} \langle F_n \rangle 
    \equiv \frac{1}{(\widehat{N}_H)^2},\\
  &&D_{H,2n_1,2n_2}\equiv \sqrt{\frac{(2n_1+1)!!(2n_2+1)!!}{(2n_1)!!(2n_2)!!}} 
         \left( \frac{2\sqrt{\nu_N\gamma}}{\nu_N+\gamma} \right)^3
         \left( \frac{\nu_N-\gamma}{\nu_N+\gamma} \right)^{n_1+n_2},
\end{eqnarray}
By combining all these formulas we have 
\begin{eqnarray}
  && M({\rm E0},0_2^+ - 0_1^+) = \frac{1}{2} \frac{{\widehat N}_H}
     {{\widehat N}_{g0}} ( D_{H,6,4} + D_{H,4,6} ) \langle R_{40}(r,\nu_N) | 
     r^2 | R_{60}(r,\nu_N) \rangle \\ 
  && \hspace{0.5cm} = \frac{1}{2} 
     \sqrt{ \frac{\langle {\cal R}^{8,J=0}_{4,4,0} \rangle}
     {\displaystyle{\left( \frac{2\sqrt{\nu_N\gamma}}{\nu_N+\gamma} \right)^6 }
     \sum_{n=5}^\infty \displaystyle{ \left( 
     \frac{\nu_N-\gamma}{\nu_N+\gamma} \right)^{2n} }
     \langle F_n \rangle} }    \nonumber \\ 
  && \hspace{1cm} \times 2 \sqrt{\frac{5!!7!!}{4!!6!!}}
     \left( \frac{2\sqrt{\nu_N\gamma}}{\nu_N+\gamma} \right)^3 
     \left( \frac{\nu_N-\gamma}{\nu_N+\gamma} \right)^5 
     \langle R_{40}(r,\nu_N) | r^2 | R_{60}(r,\nu_N) \rangle \\ 
  && \hspace{0.5cm} = \sqrt{\frac{7}{6}} 
     \sqrt{\frac{\langle F_4 \rangle}{\langle F_5 \rangle}} \xi_5 
    \langle R_{40}(r,\nu_N) | r^2 | R_{60}(r,\nu_N) \rangle, \\
  && \xi_5 \equiv \sqrt{\frac{\langle F_5 \rangle}
    {\langle F_5 \rangle + \sum_{n=6}^\infty 
    \displaystyle{ \left( \frac{\nu_N-\gamma}{\nu_N+\gamma} \right)^{2(n-5)}} 
    \langle F_n \rangle}}.   
\end{eqnarray}

\section{Estimation of $\eta_N$ and $\zeta_N$ by Ref.~\cite{suzuki}} 
\label{suzuki}

In Ref.~\cite{suzuki} the $^{16}$O states are expressed by microscopic 
$^{12}$C + $\alpha$ cluster wave functions where $^{12}$C cluster can be 
excited to its first $2^+$ and $4^+$ states.  
This coupled channel problem is solved by using the coupled channel 
OCM (orthogonality condition model).  
The wave functions $\Phi$ are expressed by the SU(3)-coupled basis of the 
$^{12}$C + $\alpha$ cluster model space, which is in the case of $J=0$ 
\begin{eqnarray}
  && \Phi = \sum_{Nq} \xi_{Nq} \Phi_{Nq}, \\ 
  && \Phi_{N,q=(\lambda,\mu)} = N_{Nq} \frac{1}{\sqrt{\mu_{Nq}}} 
     \frac{1}{\sqrt{{}_{16}C_4}} {\cal A} 
     \{ [{\cal R}_N(\Vec{r},3\nu_N) \phi({}^{12}{\rm C})]_{(\lambda,\mu)}  
    \phi(\alpha) \}, \\ 
  && \mu_{N,q=(\lambda,\mu)} = \langle [{\cal R}_N(\Vec{r},3\nu_N) 
    \phi({}^{12}{\rm C})]_{(\lambda,\mu)}  \phi(\alpha) | {\cal A} 
     \{ [{\cal R}_N(\Vec{r},3\nu_N) \phi({}^{12}{\rm C})]_{(\lambda,\mu)}  
    \phi(\alpha) \} \rangle.
\end{eqnarray}
The SU(3)-coupled basis wave functions $\Phi_{N,q=(\lambda,\mu)}$ are 
ortho-normalized and are the eigenfunctions of normalization kernel 
with eigenvalues $\mu_{N,q=(\lambda,\mu)}$.  The values of the 
expansion coefficients $\xi_{Nq}$ are given in Table I of the second 
paper of Ref.~\cite{suzuki}. 

Since the wave function of the $0_2^+$ state shows predominantly a 
$^{12}$C($0^+$) + $\alpha$ structure, we approximated in this paper 
the wave function of the $0_2^+$ state by a pure $^{12}$C($0^+$) + 
$\alpha$ structure 
\begin{eqnarray}
   \Phi(0^+_2) &=& \sum_{N \ge 6} \eta_N \Phi_N, \\ 
   \Phi_N &=& \frac{1}{\sqrt{\tau_{0,N}}} \frac{1}{\sqrt{{}_{16}C_4}} 
    {\cal A} \{ R_{N0}(r) Y_{00}(\Vec{\hat r}) \phi_{L=0}({}^{12}{\rm C}) 
    \phi(\alpha) \}. 
\end{eqnarray}
This means we adopted the following approximation for each $N$ 
\begin{equation}
   \sum_q \xi_{Nq} \Phi_{Nq} \approx \eta_N \Phi_N. 
\end{equation}
Therefore we have 
\begin{equation}
   \eta_N \approx \sqrt{\sum_q (\xi_{Nq})^2}.  
\end{equation}
The values $\sum_q (\xi_{Nq})^2$ are tabulated in Table I of the first paper 
of Ref.~\cite{suzuki}.  For $N=6$, we have $\eta_6 \approx \sqrt{0.144} = 
0.379$. 

From Table I of the second paper of Ref.~\cite{suzuki}, we have in the case 
of $N=6$ for $0_2^+$   
\begin{equation}
   \sum_q \xi_{6q} \Phi_{6q} = 0.338 \Phi_{6,(4,2)} + 0.171 \Phi_{6,(2,0)}. 
   \ \  ({\rm for}\  0_2^+) \label{eq:ocmA}
\end{equation}
We should note here that $\Phi_{6,(2,0)}$ is nothing but $\Phi_{(2,0)}^{cl}$ 
which absorbs the total monopole strength of $\Phi_{CS}$ within the 
$^{12}$C + $\alpha$ cluster model space.  On the other hand, there holds 
\begin{equation}
   \Phi_6 =  0.750 \Phi_{6,(4,2)} + 0.661 \Phi_{6,(2,0)}.  
\end{equation}
This relation is easily obtained by comparing $\langle \Phi_6 | O_M | 
\Phi_{CS} \rangle$ with $\langle \Phi_{6,(2,0)} | O_M | \Phi_{CS} \rangle$: 
\begin{eqnarray}
  && \langle \Phi_6 | O_M | \Phi_{CS} \rangle = \frac{1}{2} 
    \sqrt{ \frac{\tau_{0,4}}{\tau_{0,6}}} \langle R_{40}(r,\nu_N) | r^2 | 
    R_{60}(r,\nu_N) \rangle  
    = \frac{0.784}{\nu_N}, \\ 
  && \langle \Phi_{6,(2,0)} | O_M | \Phi_{CS} \rangle = \frac{1.19}{\nu_N}.  
\end{eqnarray}
Since the monopole operator $O_M$ does not connect $\Phi_{CS}$ with 
$\Phi_{6,(4,2)}$, we have 
\begin{eqnarray}
  \langle \Phi_{6,(2,0)} | \Phi_6 \rangle = \frac{0.784}{1.19} = 0.661. 
\end{eqnarray}
By using $\eta_6 = 0.379$ we have 
\begin{eqnarray}
 \eta_6 \Phi_6 = 0.284 \Phi_{6,(4,2)} + 0.250 \Phi_{6,(2,0)}. 
     \label{eq:0-0comp}
\end{eqnarray}
The comparison of  Eq.~(\ref{eq:ocmA}) with Eq.~(\ref{eq:0-0comp}) tells 
us that the approximation of $\sum_q \xi_{6q} \Phi_{6q}$ with 
$\eta_6 \Phi_6$ is not very good and that the monopole strength of 
 $\sum_q \xi_{6q} \Phi_{6q}$ is weaker than that of $\eta_6 \Phi_6$.  

The arguments for the $0_3^+$ state can be made completely in the same way 
as the $0_2^+$ state.  We obtain following relations 
\begin{eqnarray}
  && \zeta_6 \approx \sqrt{0.325} = 0.570, \\
  && \sum_q \xi_{6q} \Phi_{6q} = 0.471 \Phi_{6,(4,2)} + 0.321 \Phi_{6,(2,0)}, 
     \ \  ({\rm for}\  0_3^+),  \\ 
  && \Psi_6 =  0.493 \Phi_{6,(4,2)} + 0.870 \Phi_{6,(2,0)}, \\ 
  && \zeta_6 \Psi_6 = 0.281 \Phi_{6,(4,2)} + 0.496 \Phi_{6,(2,0)}. 
\end{eqnarray}
The approximation of $\sum_q \xi_{6q} \Phi_{6q}$ with $\zeta_6 \Psi_6$ is 
not also so good and the monopole strength of $\sum_q \xi_{6q} \Phi_{6q}$ 
is also weaker than that of $\zeta_6 \Psi_6$.


\newpage
\begin{table}
\caption
{Values of $\tau_{0,N}$, $\tau_{2,N}$ and $\tau_{4,N}$ which are calculated
 with the analytical expression given 
 in Refs.~\cite{suzuki,horic}.}
\label{tab:1}
\begin{center}
\begin{tabular}{cccc} 
\hline
\hline
\hspace*{10mm}$N$\hspace{10mm} & \hspace*{10mm}$\tau_{0,N}$\hspace*{10mm}  
                               & \hspace*{10mm}$\tau_{2,N}$\hspace*{10mm}  
                               & \hspace*{10mm}$\tau_{4,N}$\hspace*{10mm}  \\ 
\hline 
4  &  0.2963        &  1.4815      &  2.6667   \\ 
6  &  0.3160        &  0.7831      &  1.0243   \\ 
8  &  0.5615        &  0.7743      &  0.8773   \\ 
10 &  0.7405        &  0.8551      &  0.8985   \\ 
12 &  0.8564        &  0.9182      &  0.9374   \\ 
14 &  0.9247        &  0.9567      &  0.9654   \\ 
16 &  0.9619        &  0.9780      &  0.9818   \\ 
18 &  0.9811        &  0.9890      &  0.9907   \\ 
20 &  0.9908        &  0.9946      &  0.9953   \\ 
22 &  0.9955        &  0.9974      &  0.9977   \\ 
24 &  0.9979        &  0.9987      &  0.9989   \\ 
\hline
\hline  
\end{tabular}
\end{center}
\end{table}

\newpage
\begin{table}
\begin{center}
\caption
{$\xi_5$ and monopole matrix element $M({\rm E0},0_2^+ - 0_1^+)$ in $^{12}$C calculated 
 at several values of $\gamma$. 
$M({\rm E0},0_2^+ - 0_1^+)$ is given as $\xi_5\times 0.882/\nu_N$ ${\rm fm}^2$, 
 where the value $\nu_N=0.168$ ${\rm fm}^{-2}$ is used. 
$R_{\rm rms}(0_2^+)$ 
 is the corresponding rms radius of the Hoyle state to the adopted values 
 of $\nu_N$ and $\gamma$. 
}
\label{tab:2}
\vspace{5mm}
\begin{tabular}{cccc}
\hline\hline
\hspace*{5mm}$\gamma$ $[{\rm fm}^{-2}]$\hspace*{5mm} & \hspace*{5mm}$\xi_5$\hspace*{5mm}
 & \hspace*{5mm}$M({\rm E0}, 0_2^+ - 0_1^+)$ $[{\rm fm}^2]$\hspace*{5mm} 
 & \hspace*{5mm}$R_{\rm rms}(0_2^+)$ [fm]\hspace*{5mm} \\
\hline
$0.0238$ & $0.338$ & $1.775$ & $3.56$ \\
$0.0182$ & $0.252$ & $1.326$ & $3.79$ \\
$0.0143$ & $0.191$ & $1.004$ & $4.03$ \\
$0.0115$ & $0.147$ & $0.773$ & $4.28$ \\
\hline\hline
\end{tabular}
\end{center}
\end{table}

\newpage
\begin{table}
\begin{center}
\caption
{Dependence of the monopole matrix element in $^{12}$C on the amount of $3\alpha$-like correlation
 involved in the ground state, which is characterized by ${\widetilde \gamma}/\nu_N$. 
The monopole matrix element is given as
 $M({\rm E0},0_2^+ - 0_1^+)$=$\langle \Psi_G({\widetilde \gamma},\nu_N)
 | \sum_{i=1}^{12}(\Vec{r}_i-\Vec{r}_G)^2/2 | \Psi_H(\gamma,{\widetilde \gamma},\nu_N) \rangle$,
 where $\Psi_G({\widetilde \gamma},\nu_N)$ and $\Psi_H(\gamma,{\widetilde \gamma},\nu_N)$ 
 are the ground state and Hoyle state wave functions, respectively. 
The rms radius of the ground-state wave function $\Psi_G({\widetilde \gamma},\nu_N)$ is
 fixed to the experimental one ($2.47$~fm). Then, the ratio ${\widetilde \gamma}/\nu_N$ is
 only the parameter to describe the property of the ground state.
For a given value of ${\widetilde \gamma}/\nu_N$, the value of $\gamma$ in 
 $\Psi_H(\gamma,{\widetilde \gamma},\nu_N)$ is chosen so as to reproduce the rms radius 
 of the Hoyle state ($3.8$ fm).
See the text for details. 
}
\label{tab:3}
\vspace{5mm}
\begin{tabular}{ccc}
\hline\hline
\hspace*{5mm}${\widetilde \gamma}/\nu_N$ \hspace*{5mm} & \hspace*{5mm}$M({\rm E0}, 0_2^+ - 0_1^+)$ $[{\rm fm}^2]$ \hspace*{5mm} \\
\hline
$1.000$ & $1.326$  \\
$0.705$ & $1.810$  \\
$0.498$ & $2.473$  \\
$0.309$ & $3.597$  \\
$0.274$ & $4.035$  \\
\hline\hline
\end{tabular}
\end{center}
\end{table}

\newpage

\end{document}